\documentclass[acmtog,dvipsnames,nonacm,screen]{acmart}

\setcitestyle{square}

\usepackage{units}
\usepackage[justification=centering]{subcaption}
\usepackage{booktabs} 
\usepackage[flushleft]{threeparttable}
\usepackage{bbding}  
\usepackage{amssymb}
\usepackage{wrapfig}
\usepackage{makecell}
\usepackage{multirow}
\usepackage[]{paralist}

\usepackage{enumitem}
\usepackage{arydshln} 
\usepackage{tikz}
\usetikzlibrary{spy,calc,shapes,arrows,positioning}
\usepackage{gensymb}
\usepackage[percent]{overpic}
\usepackage{paralist}
\usepackage{multirow}
\usepackage{pifont}
\usepackage{amssymb}
\usepackage{pgfplots}
\usepackage{wrapfig}

\newcommand{\figref}[1]{Fig.\ \ref{#1}}
\newcommand{\secref}[1]{Sec.\ \ref{#1}}
\newcommand{\equref}[1]{Eq.\ \eqref{#1}}

\newcommand{\changed}{}
\newcommand{\chch}{}

\newcommand{\transposeSign}{{\!\top\!}}
\newcommand{\tr}{\transposeSign}

\usepackage{dsfont}

\newcommand{\RR}{\mathcal{R}}
\newcommand{\DD}{\mathcal{D}}

\newcommand{\GG}{\mathcal{G}} 
\newcommand{\LL}{\mathcal{L}} 
\newcommand{\TT}{\mathcal{T}} 
\newcommand{\CC}{\mathcal{C}} 

\newcommand{\Image}{\mathbb{I}}

\renewcommand{\vec}[1]{\ensuremath{\mathbf{#1}}}

\newcommand{\II}{\vec{I}}
\newcommand{\JJ}{\vec{J}}
\newcommand{\JJinv}{\vec{J}^{-1}}
\newcommand{\VV}{\vec{V}}
\newcommand{\pp}{\vec{p}}
\newcommand{\xx}{\vec{x}}
\newcommand{\ww}{\vec{w}}

\newcommand{\nn}{\vec{n}}
\newcommand{\WW}{\vec{W}}

\newcommand{\qq}{\vec{q}}

\makeatletter
\DeclareRobustCommand\onedot{\futurelet\@let@token\@onedot}
\def\@onedot{\ifx\@let@token.\else.\null\fi\xspace}

\newcommand{\xmark}{\ding{55}}%
\newcommand{\cmark}{\ding{51}}%

\makeatother

\makeatletter
\renewenvironment{cases}[1][l]{\matrix@check\cases\env@cases{#1}}{\endarray\right.}
\def\env@cases#1{%
	\let\@ifnextchar\new@ifnextchar
	\left\lbrace\def\arraystretch{1.2}%
	\array{@{}#1@{\quad}l@{}}}
\makeatother

\pgfplotsset{
	every first x axis line/.style={},
	every first y axis line/.style={},
	every first z axis line/.style={},
	every second x axis line/.style={},
	every second y axis line/.style={},
	every second z axis line/.style={},
	first x axis line style/.style={/pgfplots/every first x axis line/.append style={#1}},
	first y axis line style/.style={/pgfplots/every first y axis line/.append style={#1}},
	first z axis line style/.style={/pgfplots/every first z axis line/.append style={#1}},
	second x axis line style/.style={/pgfplots/every second x axis line/.append style={#1}},
	second y axis line style/.style={/pgfplots/every second y axis line/.append style={#1}},
	second z axis line style/.style={/pgfplots/every second z axis line/.append style={#1}}
}
\makeatletter
\def\pgfplots@drawaxis@outerlines@separate@onorientedsurf#1#2{%
	\if2\csname pgfplots@#1axislinesnum\endcsname
	\else
	\scope[/pgfplots/every outer #1 axis line,
	#1discont,decoration={pre length=\csname #1disstart\endcsname, post length=\csname #1disend\endcsname}]
	\pgfplots@ifaxisline@B@onorientedsurf@should@be@drawn{0}{%
		\draw [/pgfplots/every first #1 axis line] decorate {
			\pgfextra
			\pgfplotspointonorientedsurfaceabsetupfor{#2}{#1}{\pgfplotspointonorientedsurfaceN}%
			\pgfplots@drawgridlines@onorientedsurf@fromto{\csname pgfplots@#2min\endcsname}%
			\endpgfextra 
		};
	}{}%
	\pgfplots@ifaxisline@B@onorientedsurf@should@be@drawn{1}{%
		\draw [/pgfplots/every second #1 axis line] decorate {
			\pgfextra
			\pgfplotspointonorientedsurfaceabsetupfor{#2}{#1}{\pgfplotspointonorientedsurfaceN}%
			\pgfplots@drawgridlines@onorientedsurf@fromto{\csname pgfplots@#2max\endcsname}%
			\endpgfextra 
		};
	}{}%
	\endscope
	\fi
}%
\makeatother
\usepackage{hyperref}
\hypersetup{urlcolor=blue}

\acmJournal{TOG}

\setcopyright{rightsretained}

\acmYear{2019}\acmVolume{38}\acmNumber{6}\acmArticle{230}\acmMonth{11} \acmDOI{10.1145/3355089.3356513}
\begin{document}
	
	\title{Differentiable Surface Splatting for Point-based Geometry Processing}
	\author{Wang Yifan}
	\email{yifan.wang@inf.ethz.ch}
	\affiliation{%
		\institution{ETH Zurich}
		\country{Switzerland}
	}
	\author{Felice Serena}
	\email{fserena@student.ethz.ch}
	\affiliation{%
		\institution{ETH Zurich}
		\country{Switzerland}
	}
	
	\author{Shihao Wu}
	\email{shihao.wu@inf.ethz.ch}
	\affiliation{%
		\institution{ETH Zurich}
		\country{Switzerland}
	}

	\author{Cengiz {\"O}ztireli}
	\email{cengiz.oztireli@disneyresearch.com}
	\affiliation{%
		\institution{Disney Research Zurich}
		\country{Switzerland}
	}
	
	\author{Olga Sorkine-Hornung}
	\email{sorkine@inf.ethz.ch}
	\affiliation{%
		\institution{ETH Zurich}
		\country{Switzerland}
	}
	\begin{abstract}
		We propose Differentiable Surface Splatting (DSS), a high-fidelity differentiable renderer for point clouds. Gradients for point locations and normals are carefully designed to handle discontinuities of the rendering function. Regularization terms are introduced to ensure uniform distribution of the points on the underlying surface. We demonstrate applications of DSS to inverse rendering for geometry synthesis and denoising, where large scale topological changes, as well as small scale detail modifications, are accurately and robustly handled without requiring explicit connectivity, outperforming state-of-the-art techniques.
The data and code are at \url{https://github.com/yifita/DSS}.
	\end{abstract}

	
	\begin{teaserfigure}
		\centering
		\includegraphics[width=\linewidth]{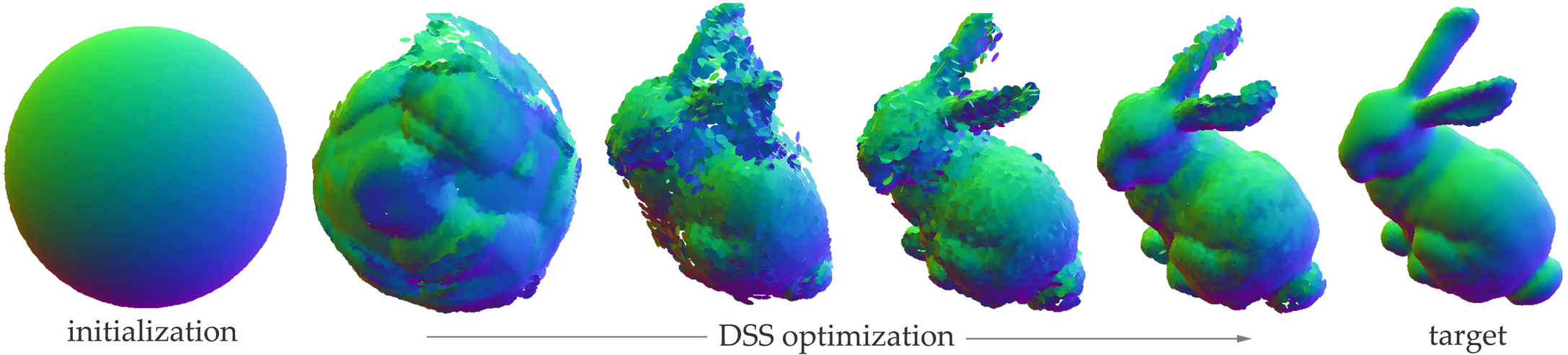}
		\caption{Using our differentiable point-based renderer, scene content can be optimized to match target rendering. Here, the positions and normals of points are optimized in order to reproduce the reference rendering of the Stanford bunny. It successfully deforms a sphere to a target bunny model, capturing both large scale and fine-scale structures. From left to right are the input points, the results of iteration 18, 57, 198, 300, and the target. }
		\label{fig:teaser}
	\end{teaserfigure}

	\maketitle
	
\section{Introduction}

Differentiable processing of scene-level information in the image formation process is emerging as a fundamental component for both 3D scene and 2D image and video modeling. The challenge of developing a differentiable renderer lies at the intersection of computer graphics, vision, and machine learning, and has recently attracted a lot of attention from all communities due to its potential to revolutionize digital visual data processing and high relevance for a wide range of applications, especially when combined with the contemporary neural network architectures~\cite{loper2014opendr, kato2018neural,liu2018paparazzi,yao20183d,petersen2019pix2vex}. 

A differentiable renderer (DR) $\mathcal{R}$ takes scene-level information $\theta$ such as 3D scene geometry, lighting, material and camera position as input, and outputs a synthesized image  $\Image = \RR(\theta)$. Any changes in the image $\Image$ can thus be propagated to the parameters $\theta$, allowing for image-based manipulation of the scene. Assuming a differentiable loss function $\LL(\Image) = \LL(\RR(\theta))$ on a rendered image $\Image$, we can update the parameters $\theta$ with the gradient $\frac{\partial \LL}{\partial \Image} \frac{\partial \Image}{\partial \theta}$. This view provides a generic and powerful shape-from-rendering framework where we can exploit vast image datasets available, deep learning architectures and computational frameworks, as well as pre-trained models. The challenge, however, is being able to compute the gradient $\frac{\partial \Image}{\partial \theta}$ in the renderer. 

Existing DR methods can be classified into three categories based on their geometric representation: voxel-based~\cite{nguyen2018rendernet,tulsiani2017multi,liu2017material}, mesh-based~\cite{loper2014opendr,kato2018neural, liu2018paparazzi}, and point-based~\cite{insafutdinov2018unsupervised,lin2018learning,roveri2018pointpronets,rajeswar2018pix2scene}. Voxel-based methods work on volumetric data and thus come with high memory requirements even for relatively coarse geometries. Mesh-based DRs solve this problem by exploiting the sparseness of the underlying geometry in the 3D space. However, they are bound by the mesh structure with limited room for global and topological changes, as connectivity is not differentiable. Equally importantly, acquired 3D data typically comes in an unstructured representation that needs to be converted into a mesh form, which is itself a challenging and error-prone operation. \emph{Point-based} DRs circumvent these problems by directly operating on point samples of the geometry, leading to flexible and efficient processing. However, existing point-based DRs use simple rasterization techniques such as forward-projection or depth maps, and thus come with well-known deficiencies in point cloud processing when capturing fine geometric details, dealing with gaps and occlusions between near-by points, and forming a continuous surface.

In this paper, we introduce Differentiable Surface Splatting (DSS), the first \emph{high fidelity point based differentiable renderer}. We utilize ideas from surface splatting~\cite{zwicker2001surface}, where each point is represented as a disk or ellipse in the object space, which is projected onto the screen space to form a splat. The splats are then interpolated to encourage hole-free and antialiased renderings. For inverse rendering, we carefully design gradients with respect to point locations and normals by taking each forward operation apart and utilizing domain knowledge. In particular, we introduce regularization terms for the gradients to carefully drive the algorithms towards the most plausible point configuration. There are infinitely many ways splats can form a given image due to the high degree of freedom of point locations and normals. Our inverse pass ensures that points stay on local geometric structures with uniform distribution. 

We apply DSS to render multi-view color images as well as auxiliary maps from a given scene. We process the rendered images with state-of-the-art techniques and show that this leads to high-quality geometries when propagated utilizing DSS. Experiments show that DSS yields significantly better results compared to previous DR methods, especially for substantial topological changes and geometric detail preservation. We focus on the particularly important application of point cloud denoising. The implementation of DSS, as well as our experiments, will be available upon publication.

	\section{Related work}
\label{sec:related}

In this section we provide some background and review the state of the art in differentiable rendering and point based processing.

\begin{table*}[t]
\begin{tabular}{l c cccc cc}
    \toprule
    method & objective & position update & depth update & normal update & occlusion & silhouette change & topology change \\
    \midrule
    OpenDR    & mesh  & \cmark & \xmark & via position change & \xmark & \cmark & \xmark\\
    NMR            & mesh & \cmark      &     \xmark        & via position change & \xmark &          \cmark &   \xmark              \\
    Paparazzi     & mesh &  limited   &   limited   & via position change & \xmark &     \xmark                   &  \xmark               \\
    Soft Rasterizer & mesh & \cmark  & \cmark    & via position change & \cmark&        \cmark  &    \xmark              \\
    Pix2Vex        & mesh   & \cmark     &    \cmark    & via position change & \cmark & \cmark        & \xmark                 \\
    Ours        & points & \cmark      & \cmark    &     \cmark    & \cmark& \cmark        &     \cmark      \\
    \bottomrule
\end{tabular}
\caption{Comparison of generic differential renderers. By design, OpenDR~\cite{loper2014opendr} and NMR~\cite{kato2018neural} do not propagate gradients to depth; Paparazzi~\cite{liu2018paparazzi} has limitation in updating the vertex positions in directions orthogonal their face normals, thus can not alter the silhouette of shapes; Soft Rasterizer~\cite{liu2019soft} and Pix2Vex~\cite{petersen2019pix2vex} can pass gradient to occluded vertices, through blurred edges and transparent faces. All mesh renderers do not consider the normal field directly and cannot modify mesh topology. Our method uses a point cloud representation, updates point position and normals jointly, considers the occluded points and visibility changes and enables large deformation including topology changes.}
\label{tab:compare_DRs}
\end{table*}

\subsection{Differentiable rendering}
\label{sec:related:DR}

We first discuss general DR frameworks, followed by DRs for specific purposes.  

Loper and Black~\shortcite{loper2014opendr} developped a differentiable renderer framework called OpenDR that approximates a primary renderer and computes the gradients via automatic differentiation.
Neural mesh renderer (NMR)~\cite{kato2018neural} approximates the backward gradient for the rasterization operation using a handcrafted function for visibility changes.
Liu et al.~\shortcite{liu2018paparazzi} propose Paparazzi, an analytic DR for mesh geometry processing using image filters. 
In concurrent work, Petersen et al.~\shortcite{petersen2019pix2vex} present \emph{Pix2Vex}, a $C^{\infty}$  differentiable renderer via soft blending schemes of nearby triangles, and Liu et al.~\shortcite{liu2019soft} introduce \emph{Soft Rasterizer}, which renders and aggregates the probabilistic maps of mesh triangles, allowing flowing gradients from the rendered pixels to the occluded and far-range vertices. All these generic DR frameworks rely on mesh representation of the scene geometry. We summarize the properties of these renderers in Table~\ref{tab:compare_DRs} and discuss them in greater detail in \secref{sec:backward}.

Numerous recent works employed DR for learning based 3D vision tasks, such as single view image reconstruction~\cite{vogels2018denoising,yan2016perspective,pontes2017image2mesh,zhu2017rethinking}, face reconstruction \cite{richardson2017learning}, shape completion \cite{hu2019render4completion}, and image synthesis \cite{sitzmann2018deepvoxels}. To describe a few, Pix2Scene~\cite{rajeswar2018pix2scene} uses a point based DR to learn implicit 3D representations from images. However, Pix2Scene renders one surfel for each pixel and does not use screen space blending.  Nguyen-Phuoc et al.~\shortcite{nguyen2018rendernet} and Insafutdinov and  Dosovitskiy~\shortcite{insafutdinov2018unsupervised} propose neural DRs using a volumetric shape representation, but the resolution is limited in practice. Li et al.~\shortcite{li2018differentiable} and Azinovi\'{c} et al.~\shortcite{azinovic2019inverse} introduce a differentiable ray tracer to implement the differentiability of physics based rendering effects, handling e.g.\ camera position, lighting and texture. {\changed While DSS could be extended and adapted to the above applications, in this paper, we demonstrate its power in shape editing, filtering, and reconstruction.}

A number of works render depth maps of point sets~\cite{lin2018learning,insafutdinov2018unsupervised,roveri2018network} for point cloud classification or generation.
These renderers do not define proper gradients for updating point positions or normals, thus they are commonly applied as an add-on layer behind a point processing network, to provide 2D supervision. 
Typically, their gradients are defined either only for depth values \cite{lin2018learning}, or within a small local neighborhood around each point. 
Such gradients are not sufficient to alter the shape of a point cloud, as we show in a pseudo point renderer in \figref{fig:pseudo}.

The differentiable rendering is also relates to shape-from-shading techniques~\cite{langguth2016shading,shi2017learning,maier2017intrinsic3d,sengupta2018sfsnet} that extract shading and albedo information for geometry processing and surface reconstruction. However, the framework proposed in this paper can be used seamlessly with contemporary deep neural networks, opening a variety of new applications.

\subsection{Point-based geometry processing and rendering}
With the proliferation of 3D scanners and depth cameras, the capture and processing 3D point clouds is becoming commonplace. 
The noise, outliers, incompleteness and misalignments persisting in the raw data pose significant challenges for point cloud filtering, editing, and surface reconstruction \cite{berger2017survey}.

Early optimization based point set processing methods rely on shape priors. Alexa and colleagues~\shortcite{alexa2003computing} introduce the moving least squares (MLS) surface model, assuming a smooth underlying surface. Aiming to preserve sharp edges, \"{O}ztireli et al.~\shortcite{oztireli2009feature} propose the robust implicit moving least squares (RIMLS) surface model. Huang et al.~\shortcite{EAR2013} employ an anisotropic \emph{weighted locally optimal projection} (WLOP) operator~\cite{Lipman2007lop, Huang2009wlop} and a progressive edge aware resampling (EAR) procedure to consolidate noisy input. Lu et al.~\shortcite{lu2018gpf} formulate WLOP with a Gaussian mixture model and use point-to-plane distance for point set processing (GPF). These methods depend on the fitting of local geometry, e.g.\ normal estimation, and struggle with reconstructing multi-scale structures from noisy input.

Advanced learning-based methods for point set processing are currently emerging, encouraged by the success of deep learning. 
Based on PointNet \cite{qi2017pointnet}, PCPNET \cite{guerrero2018pcpnet} and PointCleanNet \cite{rakotosaona2019pointcleannet} estimate local shape properties from noisy and outlier-ridden point sets;  EC-Net~\cite{yu2018ec} learns point cloud consolidation and restoration of sharp features by minimizing a point-to-edge distance, but it requires edge annotation for the training data. 
Hermosilla et al.~\shortcite{hermosilla2019total} propose an unsupervised point cloud cleaning method based on Monte Carlo convolution~\cite{hermosilla2018mccnn}.
Roveri et al.~\shortcite{roveri2018pointpronets} present a projection based differentiable point renderer to convert  unordered 3D points to 2D height maps, enabling the use of convolutional layers for height map denoising before back-projecting the smoothed pixels to the 3D point cloud. In contrast  to the commonly used Chamfer or EMD loss \cite{fan2017point}, our DSS framework, when used as a loss function, is compatible with convolutional layers and is sensitive to the exact point distribution pattern.

Surface splatting is fundamental to our method. Splatting has been developed for simple and efficient point set rendering and processing in the early seminal point based works  \cite{pfister2000surfels,zwicker2001surface,zwicker2002pointshop,zwicker2004perspective}. Recently, point based techniques have gained much attention for their superior potential in geometric learning. To the best of our knowledge, we are the first to implement high-fidelity differentiable surface splatting.


\section{Method}
\label{sec:method}
In essence, a differentiable renderer $ \RR $ is designed to propagate image-level changes to scene-level parameters $ \theta $.
This information can be used to optimize the parameters so that the rendered image $\Image= \RR\left(\theta\right) $  matches a reference image $ \Image^* $.
Typically, $ \theta $ includes the coordinates, normals and colors of the points, camera position and orientation, as well as lighting.
Formally, this can be formulated as an optimization problem
\begin{equation}\label{eq:backward_overview}
 \theta^* = \arg\min_{\theta} \LL\left(\RR\left(\theta\right), \Image^*\right),
\end{equation}
where $ \LL $ is the image loss, measuring the distance between the rendered and reference images.

Methods to solve the optimization problem \eqref{eq:backward_overview} are commonly based on gradient descent which requires $ \RR $ to be differentiable with respect to $ \theta $.
However, gradients w.r.t.\ point coordinates $\pp$ and normals $\nn$, i.e., $ \frac{d\Image}{d \pp}$ and $ \frac{d\Image}{d\nn} $, are not defined everywhere, since $ \RR $ is a discontinuous function due to occlusion events and edges. 

The key to our method is two-fold.
First, we define a gradient $ \frac{d\Image}{d\pp}$ and $ \frac{d\Image}{d\nn}$ which enables information propagation from long-range pixels without additional hyper-parameters.
Second, to address the optimization difficulty that arises from the significant number of degrees of freedom due to the unstructured nature of points, we introduce regularization terms that contribute to obtaining clean and smooth surface points.

In this section, we first review screen space EWA (elliptical weighted average) \cite{zwicker2001surface,heckbert1989fundamentals}, which we adopt to efficiently render high-quality realistic images from point clouds.
Then we propose an occlusion-aware gradient definition for the rasterization step, which, unlike previously proposed differential mesh renderers, propagates gradients to depth and allows large deformation.
Lastly, we introduce two novel regularization terms for generating clean surface points.

\subsection{Forward pass}\label{sec:forward}
\begin{figure}\centering
\begin{subfigure}{0.58\linewidth}
\begin{overpic}[height=0.8\linewidth,tics=10]{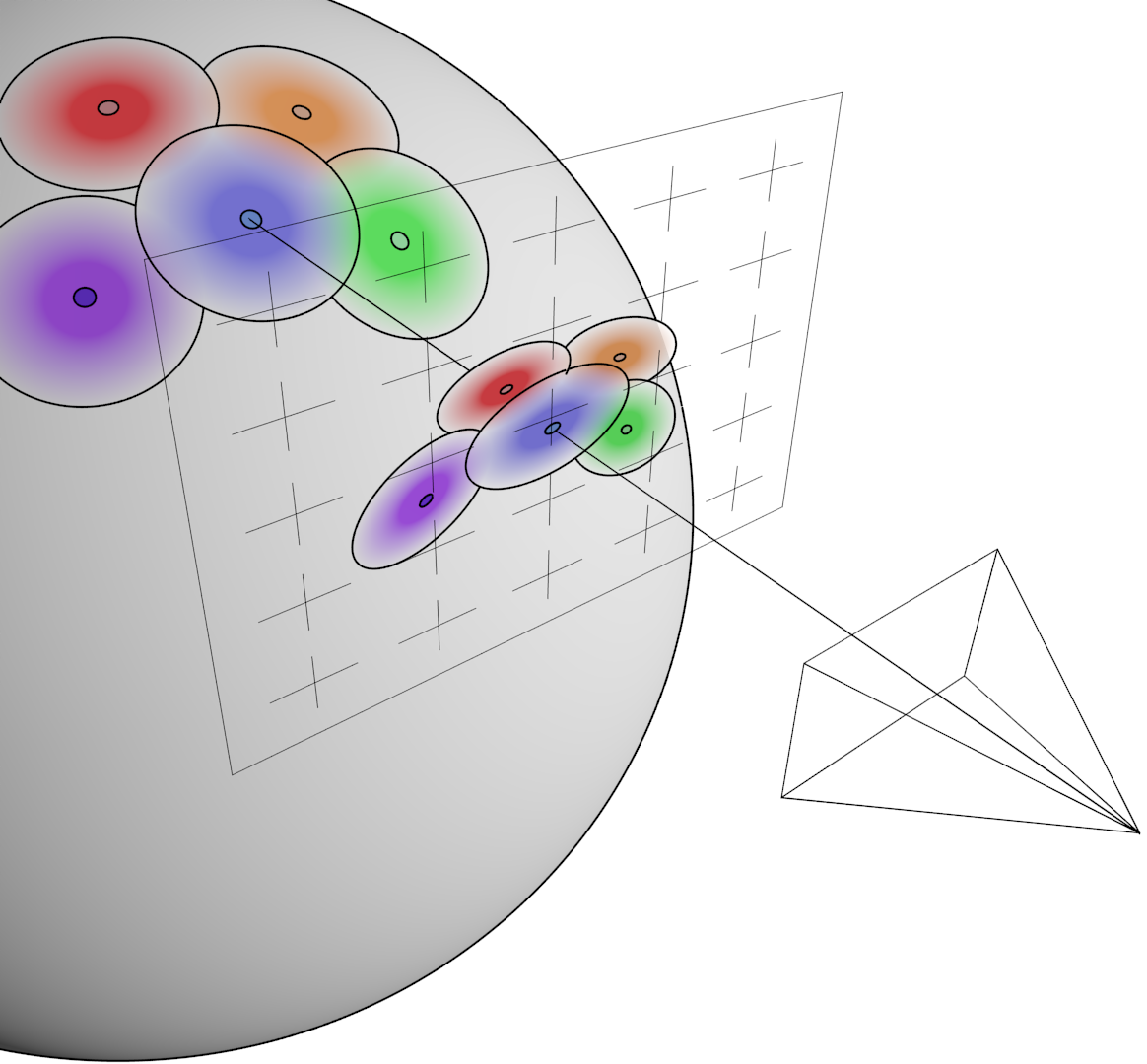}
\put (20,75) {$\displaystyle\pp_k$}
\put (42,48) {$\displaystyle\xx_k$}
\end{overpic}
\end{subfigure}
\begin{subfigure}{0.41\linewidth}
\begin{overpic}[height=0.8\linewidth]{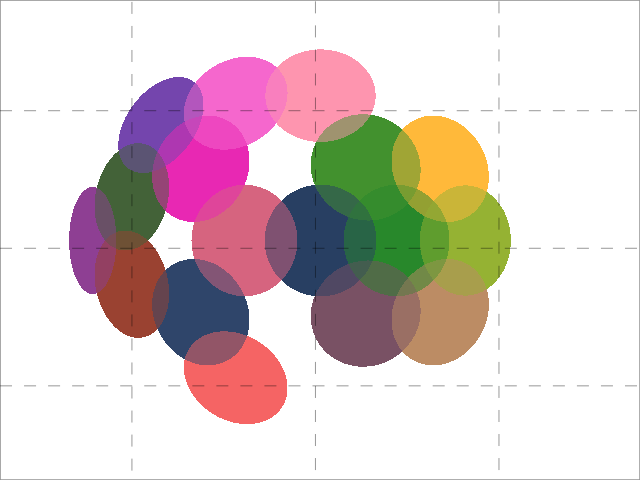}
\put (50, 30) {\color{white}$\displaystyle\xx$}
\end{overpic}
\end{subfigure}
\caption{\changed Illustration of forward splatting using EWA~\cite{zwicker2001surface}. A point in space $ \pp_{k} $ is rendered as an anisotropic ellipse centered at the projection point $ \xx_k $. The final pixel value $ \Image_\xx $ at a pixel $\xx$ in the image (shown on the right) is the normalized sum of all such ellipses overlapping at $ \xx $.}
\label{fig:spEWA}\vspace{-4ex}
\end{figure}
{\changed The forward pass refers to the generation of a 2D image from 3D scene-level information, $ \Image = \RR\left(\theta\right) $.}
Our forward pass closely follows the screen space elliptical weighted average (EWA) filtering described in \cite{zwicker2001surface}.
In the following, we briefly review the derivation of EWA filters.

In a nutshell, the idea of screen space EWA is to apply an isotropic Gaussian filter to the attribute $ \Phi $ of a point in the tangent plane (defined by the normal at that point). 
The projection onto the image plane defines elliptical Gaussians, which, after truncation to bounded support, form a disk, or splat, as shown in \figref{fig:spEWA}. 
For a point $ \pp_k $, we write the filter weight of the isotropic Gaussian at position $ \pp $ as 
\begin{align}
\GG_{\pp_k, \VV_k}\left(\pp\right) = \dfrac{1}{2\pi |\VV_k|^{\frac{1}{2}}}e^{\left(\pp-\pp_k\right)^\tr \VV_k^{-1}\left(\pp-\pp_k\right) }, \quad \VV_k = \sigma_k^2 \II,
\end{align}
where $ \sigma_k $ is the standard deviation and $ \II $ is the identity matrix.

Now we consider the projected Gaussian in screen space. 
Points $ \pp_k $ and $ \pp $ are projected to $ \xx_k $ and $ \xx $, respectively.
We write the Jacobian of this projection from the tangent plane to the image plane as $ \JJ_k $; we refer the reader to the original surface splatting paper \cite{zwicker2001surface} for the derivation of $ \JJ_k $. Then at $ \xx $, the screen space elliptical Gaussian weight is
\begin{align}
    r_{k}\left(\xx\right) &= \GG_{\VV_k}\left(\JJ_k^{-1}\left(\xx-\xx_k\right) \right) \nonumber\\ 
    &= \dfrac{1}{\left|\JJinv_k\right|}\GG_{\JJ_k\VV_{k}\JJ_k^\tr}\left(\xx-\xx_k\right).\label{eq:rk}
\end{align}
Note that $ r_{k} $ is determined by the point position $ \pp_k $ and the normal $ \nn_k $, because $ \JJ_k $ is determined by $ \pp_k $ and $ \nn_k $.

Next, a low-pass Gaussian filter with variance $ \II $ is convolved with \equref{eq:rk} in screen space.
Thus the final elliptical Gaussian is 
\begin{equation}\label{eq:filter}
\bar{\rho}_{k}\left(\xx\right) = \dfrac{1}{\left|\JJinv_k\right|}\GG_{\JJ_k\VV_{k}\JJ_k^\tr+\II}\left(\xx-\xx_k\right).
\end{equation}
In the final step, two sources of discontinuity are introduced to the fully differentiable $ \bar{\rho} $.
First, for computational reasons, we limit the elliptical Gaussians to a limited support in the image plane for all $ \xx $ outside a cutoff radius $ \CC $, i.e., $ \frac{1}{2}\xx^\tr \left(\JJ\VV_{k}\JJ^\tr+\II \right)\xx > \CC$.
Second, we set the Gaussian weights for occluded points to zero. 
Specifically, we keep a list of the maximum $ K $ (we choose $ K=5 $) closest points at each pixel position, and compute their depth difference to the front-most point, and then set the Gaussian weights to zero for points that are behind the front-most point by more than a threshold $ \TT $ (we set $ \TT = 1\% $ of the bounding box diagonal length).
These $ K $ points are cached for gradient evaluation in backward pass, as will be explained in \secref{sec:backward}.

The resulting truncated Gaussian weight, denoted as $ \rho_k $, can be formally defined as 
\begin{align}
\rho_k\left(\xx\right) = & \begin{cases}
0, &\text{if $\frac{1}{2}\xx^\tr \left(\JJ\VV_{k}\JJ^\tr+\II \right)\xx > \CC$}, \\
0, &\text{if $ \pp_{k} $ is occluded}, \\
\bar{\rho}_k, &\text{otherwise.}\label{eq:rho}
\end{cases}
\end{align}

The final pixel value is simply the normalized sum of all filtered point attributes $ \lbrace\ww_k\rbrace_{k=0}^{N} $ evaluated at the center of pixels, i.e.,
\begin{equation}
\Image_{\xx} = \dfrac{\sum_{k=0}^{N-1}\rho_{k}\left(\xx\right)\,\ww_k}{\sum_{k=0}^{N-1}\rho_k\left(\xx\right)}.\label{eq:sumI}
\end{equation}
In practice, this summation can be greatly optimized by computing the bounding box of each ellipse and only considering points whose elliptical support covers the pixel $ \xx $.

The point value $ \Phi $ can be any point attribute, e.g., albedo color, shading, depth value, normal vector, etc.
In most of our experiments, we use diffuse shading under three orthogonally positioned RGB-colored sun lights.
This way, $ \Phi $ carries strong information about point normals, and at the same time it is independent of point position (unlike with point lights), which greatly simplifies the factorization for gradient computation, as explained in~\secref{sec:backward}.

\begin{figure}
    \setlength{\tabcolsep}{1pt}
    \begin{tabular}{ccc}
        \includegraphics[width=0.3\linewidth,trim={1.5cm, 1.5cm, 1.5cm, 1.5cm}, clip]{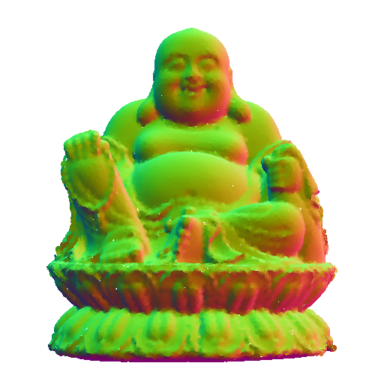}&
        \includegraphics[width=0.3\linewidth,trim={1.5cm, 1.5cm, 1.5cm, 1.5cm}, clip]{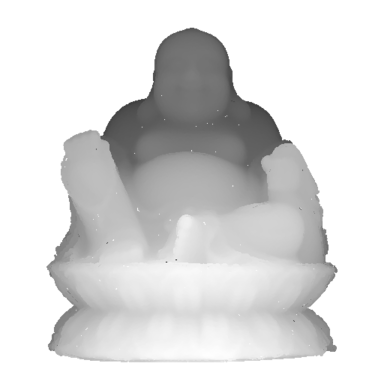}&
        \includegraphics[width=0.3\linewidth,trim={1.5cm, 1.5cm, 1.5cm, 1.5cm}, clip]{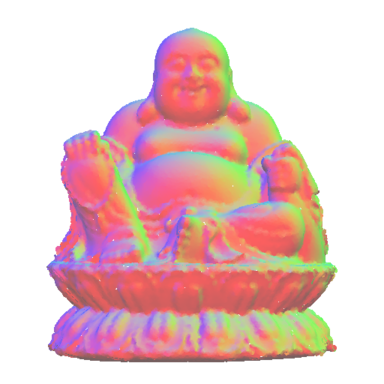}
    \end{tabular}
    \caption{Examples of images rendered using DSS. From left to right, we render the normals, inverse depth values and diffuse shading with three RGB-colored sun light sources.}
    \label{fig:render_examples}
\end{figure}

\figref{fig:render_examples} shows some examples of rendered images. 
Unlike many pseudo renderers which achieve differentiability by blurring edges and transparent surfaces, our rendered images faithfully depict the actual geometry in the scene.

\subsection{Backward pass}\label{sec:backward}


\begin{figure*}
	\begin{subfigure}[b]{0.43\textwidth}
		\centering
		\def\svgwidth{\linewidth}
		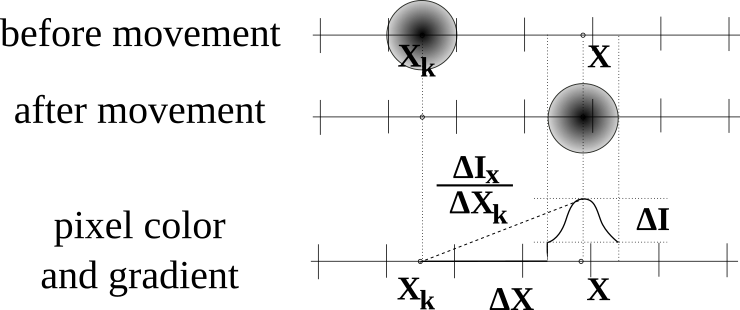\caption{The ellipse centered at $\pp_{k,0}$ is not visible at $ \xx $.}\label{fig:1D_outside}
	\end{subfigure}\hspace{0.05\textwidth}
	\begin{subfigure}[b]{0.35\textwidth}
		\def\svgwidth{\linewidth}
		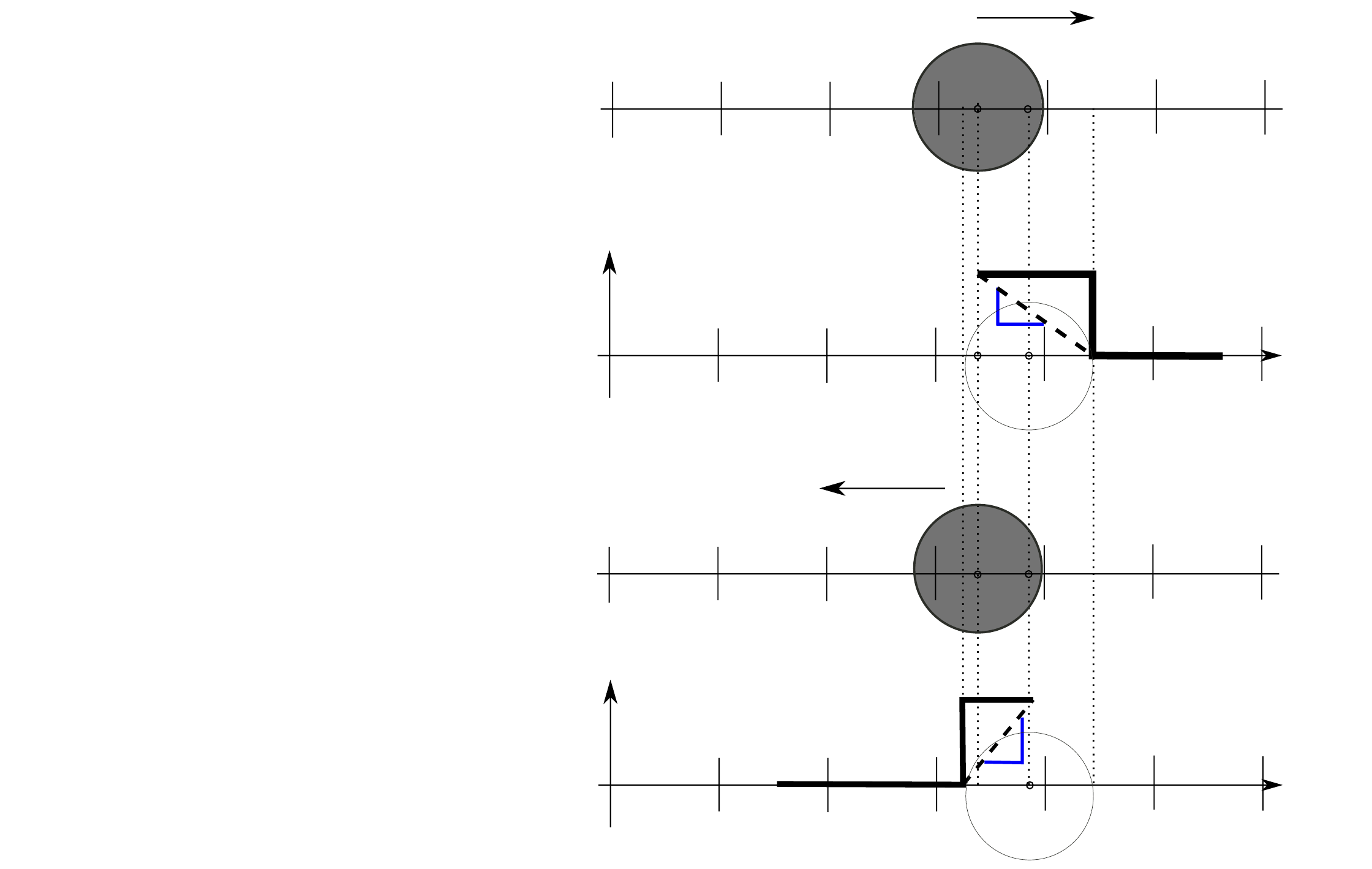\caption{The ellipse centered at $\pp_{k,0}$ is visible at $ \xx $.}\label{fig:1D_inside}
	\end{subfigure}
	\caption{\changed An illustration of the artificial gradient in two 1D scenarios: the ellipse centered at $ \pp_{k,0} $ is invisible (\figref{fig:1D_outside}) and visible (\figref{fig:1D_inside}) at pixel $ \xx $. $ \Phi_{\xx, k} $ is the pixel intensity $ \Image_{\xx}$ as a function of point position $ \pp_{k} $, $ \qq_{\xx} $ is the coordinates of the pixel  $\xx$ back-projected to world coordinates. Notice the ellipse has constant pixel intensity after normalization (\equref{eq:sumI}). We approximate the discontinuous $ \Phi_{\xx, k} $ as a linear function defined by the change of pixel intensity $ \Delta \Image_\xx $ and the movement of the $ \Delta \pp_k $ during a visibility switch. As $ \pp_k $ moves toward ($ \Delta \pp_k^{+}$) or away ($ \Delta \pp_k^{-}$) from the pixel, we obtain two different gradient values. We define the final gradient as their sum.} \label{fig:1D}
\end{figure*}
{\changed The backward pass refers to the information flow from the rendered image $ \Image = \RR\left(\theta\right)$ to the scene parameters $ \theta $ based on approximating the gradient $ \frac{d \Image}{d\theta}$.
As discussed, the key to address the discontinuity of $ \RR $ lies in the approximation of the gradient $ \frac{d\Image}{d\pp}$ and $ \frac{d\Image}{d\nn}$.}


The discontinuity of $ \RR $ is encapsulated in the truncated Gaussian weights $ \rho_k $ as described \equref{eq:rho}. 
We can factorize the discontinuous $ \rho_k $ into the fully differentiable term $ \bar{\rho}_k $ and a discontinuous visibility term $ h_\xx\in \lbrace0, 1\rbrace $, i.e., $ \rho_k = h_\xx\bar{\rho}_k $, where
$ h_\xx $ is defined as 
\begin{align}
h_\xx\left(\pp_{k}\right) = & \begin{cases}
0, & \text{if} \frac{1}{2}\xx^\tr \left(\JJ\VV_{k}\JJ^\tr+\II \right)\xx > \CC, \\
0, & \text{if $ \pp_{k} $ is occluded}, \\
1, & \text{otherwise.}\label{eq:h}
\end{cases}
\end{align}
Note that even though $ h_\xx $ is indirectly influenced by $ \nn_{k} $ through $ \JJ $, since this only impacts the visibility of a small set of pixels around the ellipse, we omit this $ \nn_k $ in this formulation.
Therefore, if we write $ \Image_\xx $ as a function of $ \ww_k $, $ \bar{\rho}_k $ and $ h_k $, then by the chain rule we have
\begin{align}
\frac{d \Image_{\xx}\left(\ww_k, \bar{\rho}_k, h_\xx\right)}{d \nn_k} = & \dfrac{\partial\Image_\xx}{\partial\ww_k}\dfrac{\partial\ww_k}{\partial\nn_k} + \dfrac{\partial\Image_\xx}{\partial\bar{\rho}_k}\dfrac{\partial\bar{\rho}_k}{\partial\nn_k},\label{eq:gradN}\\
\frac{d \Image_{\xx}\left(\ww_k, \bar{\rho}_k, h_\xx\right)}{d \pp_k} = & \dfrac{\partial\Image_\xx}{\partial\ww_k}\dfrac{\partial\ww_k}{\partial\pp_k} + \dfrac{\partial\Image_\xx}{\partial\bar{\rho}_k}\dfrac{\partial\bar{\rho}_k}{\partial\pp_k} + 
\dfrac{\partial\Image_\xx}{\partial h_\xx}\dfrac{\partial h_\xx}{\partial\pp_k},\label{eq:gradP}
\end{align}
where \equref{eq:gradN} is fully differentiable but \equref{eq:gradP} is not, as $ \frac{\partial h_\xx}{\partial\pp_k}$ is undefined at the edge of ellipses.

{\changed We focus on the partial gradient $\frac{\partial\Image_\xx}{\partial h_\xx}\frac{\partial h_\xx}{\partial \pp_{k}} $.
Denoting $ \Phi_{\xx}\left(\pp_{k}\right) = \Image_\xx\left(h_\xx\left(\pp_{k}\right)\right)$, this  gradient can be written as $ \frac{d\Phi_{\xx}}{d\pp_k} $, which describes the change of a pixel intensity $ \Image_{\xx} $ due to the visibility change of a point caused by its varying position $ \pp_{k} $.

\paragraph{1D scenario.} Let us first consider a simplified scenario where a single point only moves in 1D space.  As shown in \figref{fig:1D}, $ \Phi_{\xx} $ is generally discontinuous; it is zero almost everywhere except in a small region around $ \qq_\xx $, the coordinates of the pixel $\xx$ back-projected to world coordinates. 
Similar to NMR \cite{kato2018neural}, we approximate $ \Phi_{\xx} $ as a linear function defined by the change of point position $ \Delta \pp_{k} $ and the pixel intensity $ \Delta \Image $ before and after the visibility change.

As $ \pp_{k} $ moves toward or away from  $ \qq_\xx $, we obtain two different linear functions with gradients {\small $ \left.\frac{d\Phi_{\xx}}{d \pp_k}\right\vert_{\pp_{k,0}}^+$} and {\small$ \left.\frac{d\Phi_{\xx}}{d \pp_k}\right\vert_{\pp_{k,0}}^-,$} respectively. 
Specifically, when $ \pp_k $ is invisible at $ \xx $ (\figref{fig:1D_outside}), moving away will always induce zero gradient, while when $ \pp_k $ is visible, we obtain two gradients with opposite signs (\figref{fig:1D_inside}).
The final gradient is defined as the sum of both, i.e., 
\begin{equation}
\left.\frac{d\Phi_{x}}{d \pp_k}\right\vert_{\pp_{k,0}} = 
\begin{cases}
\frac{\Delta \Image_\xx}{\|\Delta \pp_{k}^+\|^2+\epsilon}\Delta \pp_{k}^{+}, &\text{$\pp_k$ invisible at $ \xx $}
\vspace{1.5ex}\\
\frac{\Delta \Image_\xx}{\|\Delta \pp_{k}^-\|^2+\epsilon}\Delta \pp_{k}^{-} + \frac{\Delta \Image_\xx}{\|\Delta \pp_{k}^+\|+\epsilon}\Delta \pp_{k}^{+}, &\text{otherwise,}
\end{cases}
\label{eq:gradcase}
\end{equation}
where $ \Delta \pp_{k}^- $ and $ \Delta \pp_{k}^+ $ denote the point movement toward and away from $ \xx $, starting from the current position $ \pp_{k,0} $.
The value $ \epsilon $ is a small constant (we set $ \epsilon = 10^{-5} $). 
It prevents the gradient from becoming extremely large when $ \pp_k $ is close $ \qq_{\xx} $, which would lead to overshooting, oscillation and other convergence problems.
}

\begin{figure}
\begin{subfigure}{0.32\linewidth}
\centering
\def\svgwidth{\linewidth}
\begingroup%
\makeatletter%
\providecommand\color[2][]{%
	\errmessage{(Inkscape) Color is used for the text in Inkscape, but the package 'color.sty' is not loaded}%
	\renewcommand\color[2][]{}%
}%
\providecommand\transparent[1]{%
	\errmessage{(Inkscape) Transparency is used (non-zero) for the text in Inkscape, but the package 'transparent.sty' is not loaded}%
	\renewcommand\transparent[1]{}%
}%
\providecommand\rotatebox[2]{#2}%
\newcommand*\fsize{\dimexpr\f@size pt\relax}%
\newcommand*\lineheight[1]{\fontsize{\fsize}{#1\fsize}\selectfont}%
\ifx\svgwidth\undefined%
\setlength{\unitlength}{461.98764219bp}%
\ifx\svgscale\undefined%
\relax%
\else%
\setlength{\unitlength}{\unitlength * \real{\svgscale}}%
\fi%
\else%
\setlength{\unitlength}{\svgwidth}%
\fi%
\global\let\svgwidth\undefined%
\global\let\svgscale\undefined%
\makeatother%
\begin{picture}(1,0.7945122)%
\lineheight{1}%
\setlength\tabcolsep{0pt}%
\put(0,0){\includegraphics[width=\unitlength,page=1]{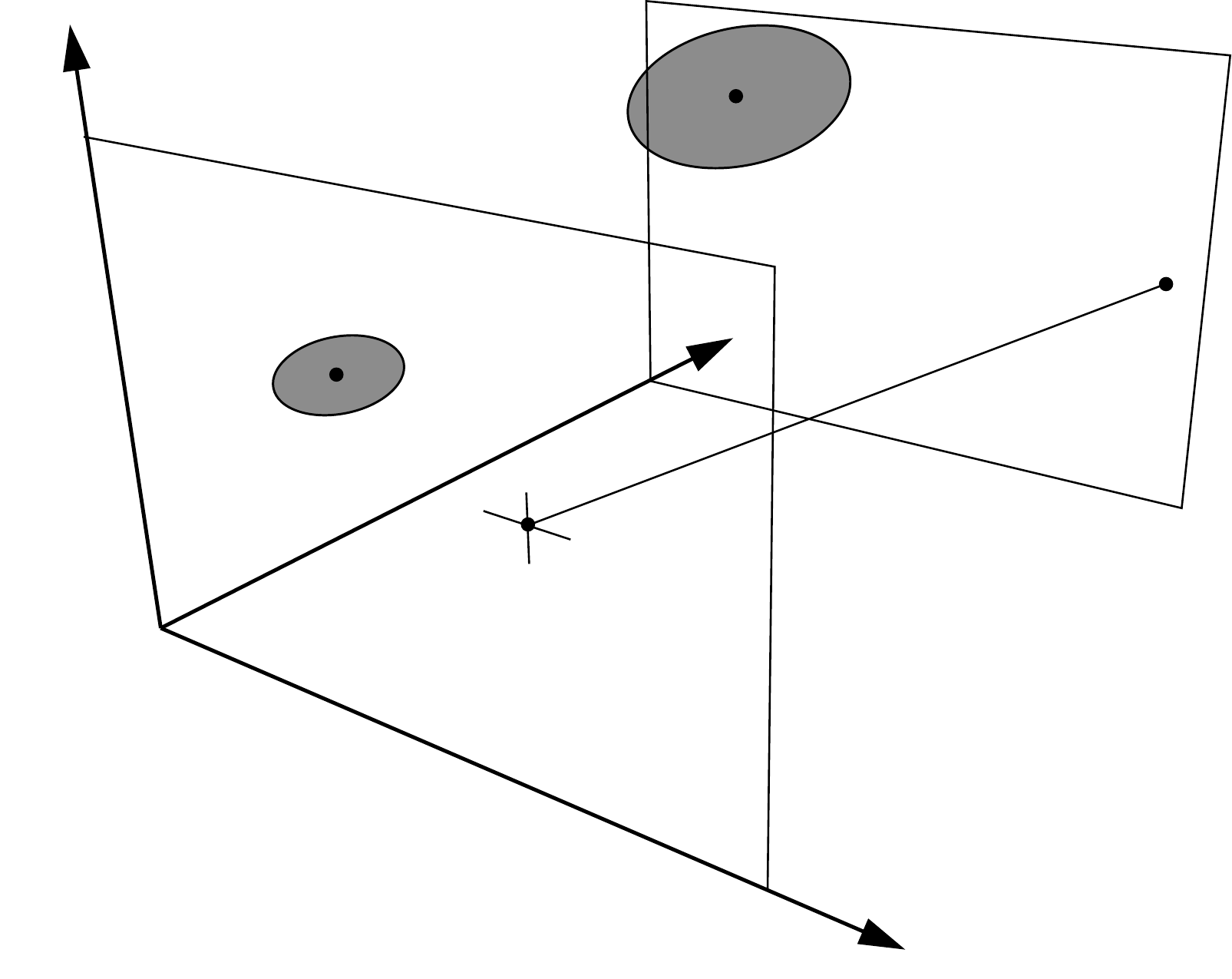}}%
\put(0.63777823,-0.02){\color[rgb]{0,0,0}\makebox(0,0)[lt]{\lineheight{1.25}\smash{\begin{tabular}[t]{l}$ x $\end{tabular}}}}%
\put(-0.00052846,0.71579788){\color[rgb]{0,0,0}\makebox(0,0)[lt]{\lineheight{1.25}\smash{\begin{tabular}[t]{l}$ y $\end{tabular}}}}%
\put(0.48706645,0.49388132){\color[rgb]{0,0,0}\makebox(0,0)[lt]{\lineheight{1.25}\smash{\begin{tabular}[t]{l}$ z $\end{tabular}}}}%
\put(0.37014597,0.32115029){\color[rgb]{0,0,0}\makebox(0,0)[lt]{\lineheight{1.25}\smash{\begin{tabular}[t]{l}$ \xx $\end{tabular}}}}%
\put(0.51345006,0.7400229){\color[rgb]{0,0,0}\makebox(0,0)[lt]{\lineheight{1.25}\smash{\begin{tabular}[t]{l}$ \pp_{k} $\end{tabular}}}}%
\put(0.90112395,0.61525294){\color[rgb]{0,0,0}\makebox(0,0)[lt]{\lineheight{1.25}\smash{\begin{tabular}[t]{l}$ \qq_{\xx} $\end{tabular}}}}%
\put(0.18218985,0.5342152){\color[rgb]{0,0,0}\makebox(0,0)[lt]{\lineheight{1.25}\smash{\begin{tabular}[t]{l}$ \xx_k $\end{tabular}}}}%
\end{picture}%
\endgroup%

\caption{}\label{fig:outside}
\end{subfigure}\hfill
\begin{subfigure}{0.32\linewidth}
\centering
\def\svgwidth{\linewidth}
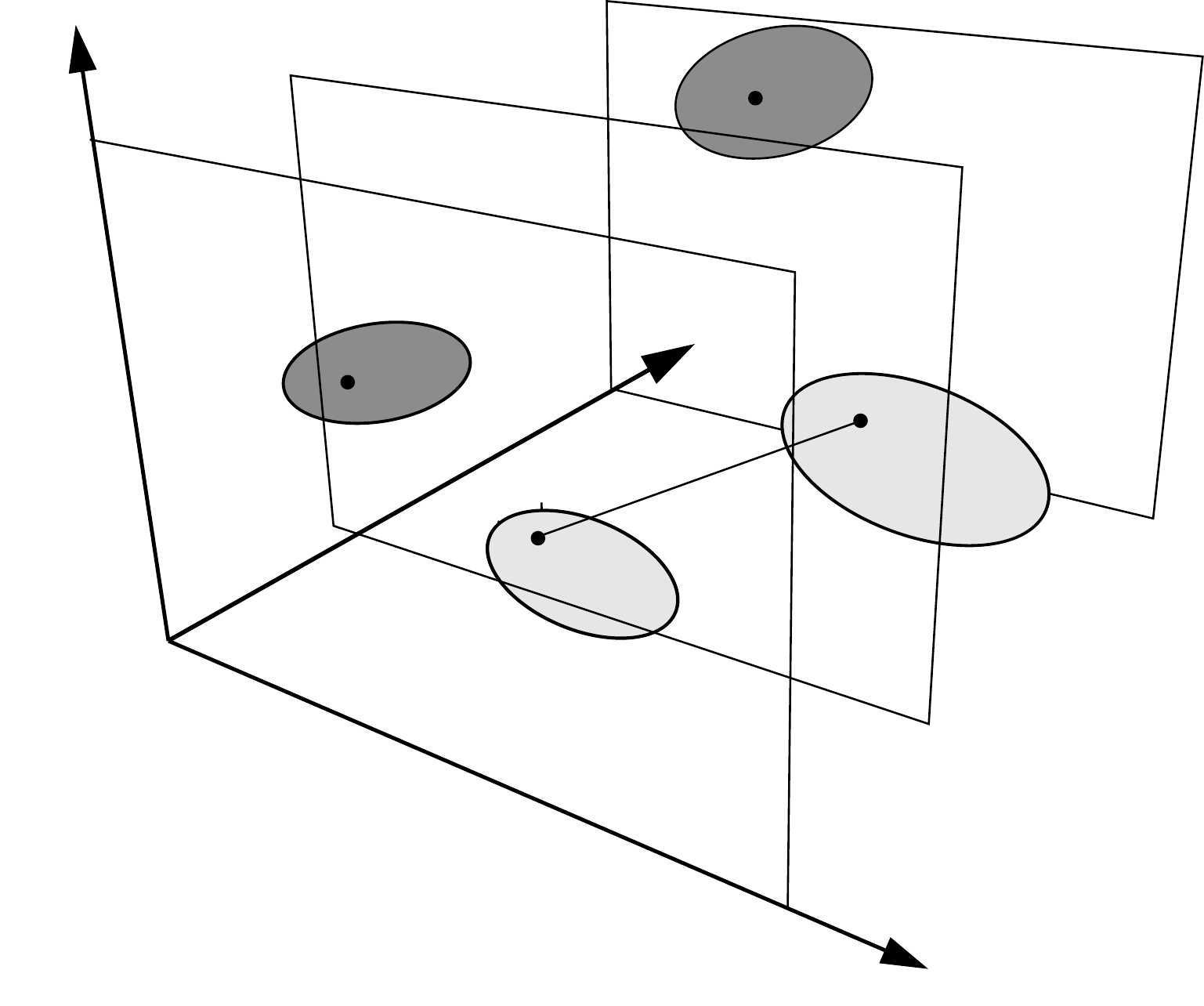
\caption{}\label{fig:outside_front}
\end{subfigure}\hfill
\begin{subfigure}{0.32\linewidth}
\def\svgwidth{\linewidth}
\begingroup%
\makeatletter%
\providecommand\color[2][]{%
	\errmessage{(Inkscape) Color is used for the text in Inkscape, but the package 'color.sty' is not loaded}%
	\renewcommand\color[2][]{}%
}%
\providecommand\transparent[1]{%
	\errmessage{(Inkscape) Transparency is used (non-zero) for the text in Inkscape, but the package 'transparent.sty' is not loaded}%
	\renewcommand\transparent[1]{}%
}%
\providecommand\rotatebox[2]{#2}%
\newcommand*\fsize{\dimexpr\f@size pt\relax}%
\newcommand*\lineheight[1]{\fontsize{\fsize}{#1\fsize}\selectfont}%
\ifx\svgwidth\undefined%
\setlength{\unitlength}{473.27452304bp}%
\ifx\svgscale\undefined%
\relax%
\else%
\setlength{\unitlength}{\unitlength * \real{\svgscale}}%
\fi%
\else%
\setlength{\unitlength}{\svgwidth}%
\fi%
\global\let\svgwidth\undefined%
\global\let\svgscale\undefined%
\makeatother%
\begin{picture}(1,0.85757649)%
\lineheight{1}%
\setlength\tabcolsep{0pt}%
\put(0,0){\includegraphics[width=\unitlength,page=1]{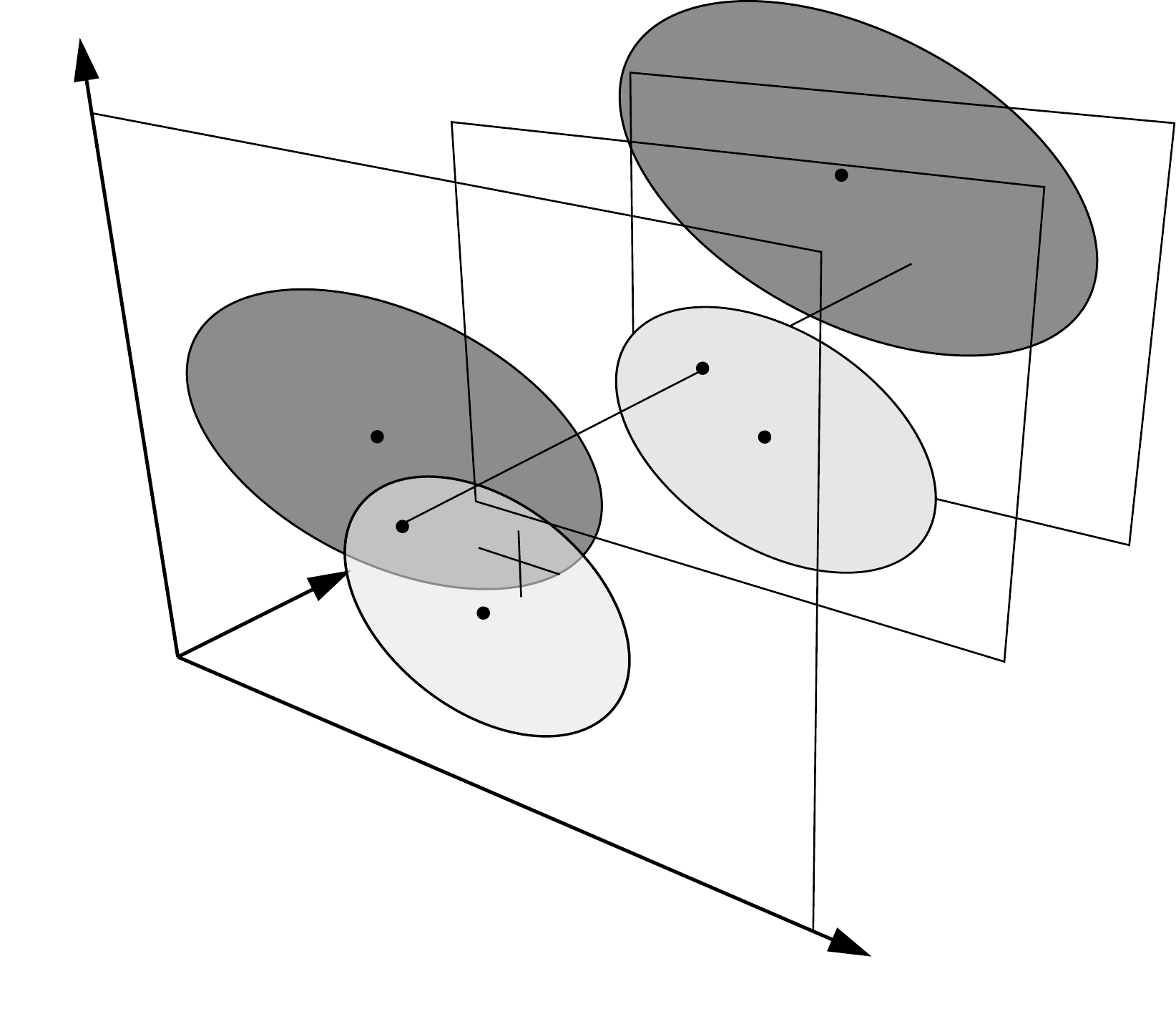}}%
\put(0.64437955,-0){\color[rgb]{0,0,0}\makebox(0,0)[lt]{\lineheight{1.25}\smash{\begin{tabular}[t]{l}$ x $\end{tabular}}}}%
\put(-0.00116602,0.74307527){\color[rgb]{0,0,0}\makebox(0,0)[lt]{\lineheight{1.25}\smash{\begin{tabular}[t]{l}$ y $\end{tabular}}}}%
\put(0.16985425,0.34545882){\color[rgb]{0,0,0}\makebox(0,0)[lt]{\lineheight{1.25}\smash{\begin{tabular}[t]{l}$ z $\end{tabular}}}}%
\put(0.27233607,0.40470424){\color[rgb]{0,0,0}\makebox(0,0)[lt]{\lineheight{1.25}\smash{\begin{tabular}[t]{l}$ \xx $\end{tabular}}}}%
\put(0.66439976,0.4993554){\color[rgb]{0,0,0}\makebox(0,0)[lt]{\lineheight{1.25}\smash{\begin{tabular}[t]{l}$ \pp_{k} $\end{tabular}}}}%
\put(0.47875744,0.5404197){\color[rgb]{0,0,0}\makebox(0,0)[lt]{\lineheight{1.25}\smash{\begin{tabular}[t]{l}$ \qq_\xx $\end{tabular}}}}%
\put(0.42712212,0.29083728){\color[rgb]{0,0,0}\makebox(0,0)[lt]{\lineheight{1.25}\smash{\begin{tabular}[t]{l}$ \xx_k $\end{tabular}}}}%
\end{picture}%
\endgroup%

\caption{}\label{fig:inside}
\end{subfigure}
\caption{\changed Illustration of the 3 cases for evaluating \equref{eq:gradcase} for 3D point clouds.}\label{fig:gradient}
\end{figure}

{\changed \paragraph{3D cases.} Extending the single point 1D-scenario to a point cloud in 3D requires evaluating $ \Delta \Image$ and $ \Delta \pp$ with care. As depicted in \figref{fig:gradient}, the following cases are considered:
\begin{inparaenum}[(a)]
\item $ \pp_{k} $ is not visible at $ \xx $ and $ \xx $ is not rendered by any other ellipses \emph{in front of} $ \pp_{k} $; \label{enum:4}
\item $ \pp_{k} $ is not visible at $ \xx $ and $ \xx $ is rendered by other ellipses \emph{in front of} $ \pp_k $; \label{enum:1}
\item $ \pp_{k} $ is visible at $ \xx $.\label{enum:2}
\end{inparaenum} 

For \eqref{enum:4} and \eqref{enum:2}, we only need to compute the gradient in screen space, whereas for \eqref{enum:1}, $ \pp_{k} $ must move forward in order to become visible, resulting in a negative depth gradient.
Furthermore, for \eqref{enum:4} and \eqref{enum:1} we evaluate the new $ \Image_{\xx} $ using \equref{eq:sumI}, \emph{adding} the contribution from $ \pp_{k} $, while for \eqref{enum:2} we need to \emph{subtract} the contribution of $ \pp_{k} ,$ which may include previously occluded ellipses into \equref{eq:sumI}.
For this purpose, as mentioned in \ref{sec:forward}, we cache an ordered list of the top-$K$ (we choose $ K $=5) closest ellipses that can be projected onto each pixel and save their $ \rho $, $ \Phi $ and depth values during the forward pass.
The value of $ K $ is related to the merging threshold $ \TT $, and as $ \TT $ is typically small, we find $ K=5 $ is sufficient even for dense point clouds.

Finally, similar to NMR, when evaluating \equref{eq:gradcase} for the optimization problem \eqref{eq:backward_overview}, we set the gradient to zero if the change of pixel intensity cannot reduce the image loss $ \LL $, i.e.,
\begin{equation}
\left.\frac{d\Phi_{x}}{d \pp_k}\right\vert_{\pp_k=\pp_{k,0}}=0  \quad\text{if}\quad  \frac{d \LL}{d \Image_\xx} \Delta\II_\xx  >= 0\label{eq:filteredG}.
\end{equation}
}

\begin{figure}\scriptsize\centering
\setlength{\tabcolsep}{1pt}
\begin{tabular}{ccccccc}
\rotatebox[origin=l]{90}{\hspace{-.5ex}RBF $ \sigma_k=5 $}&
\fbox{\includegraphics[width=0.12\linewidth]{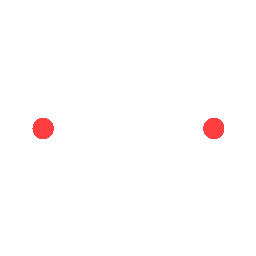}}&
\fbox{\includegraphics[width=0.12\linewidth]{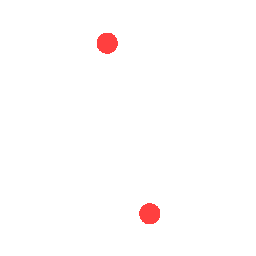}}&
\fbox{\includegraphics[width=0.12\linewidth]{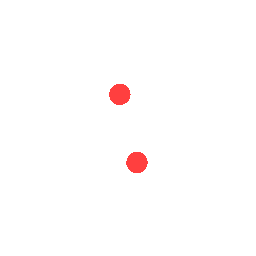}}&
\fbox{\includegraphics[width=0.12\linewidth]{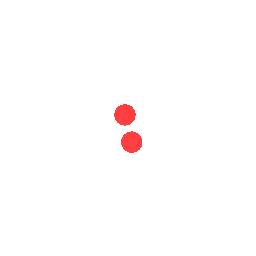}}&
\fbox{\includegraphics[width=0.12\linewidth]{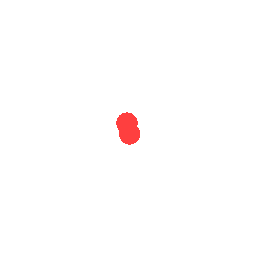}}&
\fbox{\includegraphics[width=0.12\linewidth]{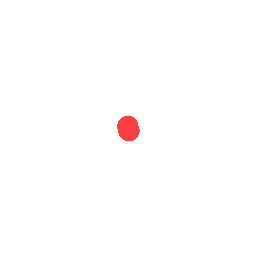}}\\
\rotatebox[origin=l]{90}{\hspace{1.8ex}Ours}&
\fbox{\includegraphics[width=0.12\linewidth]{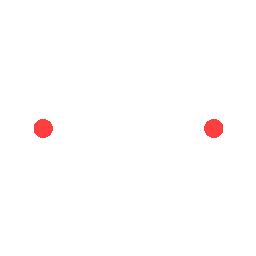}}&
\fbox{\includegraphics[width=0.12\linewidth]{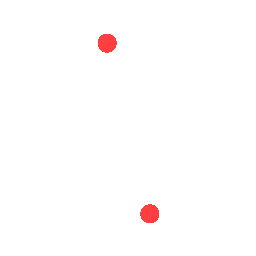}}&
\fbox{\includegraphics[width=0.12\linewidth]{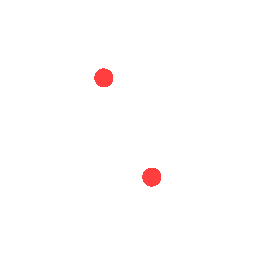}}&
\fbox{\includegraphics[width=0.12\linewidth]{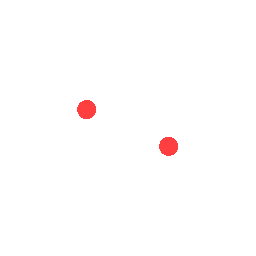}}&
\fbox{\includegraphics[width=0.12\linewidth]{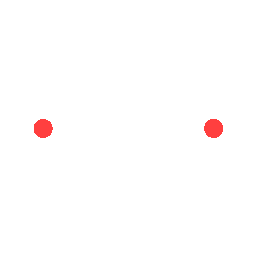}}&
\fbox{\includegraphics[width=0.12\linewidth]{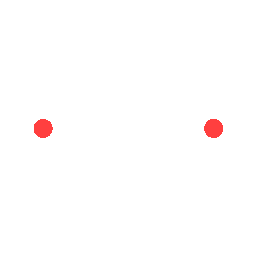}}\\
& target & initialization & step 12 & step 24 & step 36 & step 48
\end{tabular}
\caption{\chch Optimization progress using our gradient approximation and RBF-derived gradient. The RBF-derived gradient is prone to local minima when optimizing for multiple points.}\label{fig:twopoints}\vspace{-3ex}
\end{figure}

\paragraph{Comparison to other differentiable renderers} 
A few differential renderers have been proposed for meshes.
In Paparazzi~\cite{liu2018paparazzi}, the rendering function is simplified enough such that the gradients can be computed analytically, which is prohibitive for silhouette change where handling significant occlusion events is required. 
{\changed OpenDR~\cite{loper2014opendr} computes gradients only in screen space from a small set of pixels near the boundary, which is conceptually less accurate than our definition. 
SoftRasterizer \cite{liu2019soft} alters the forward rendering to make the rasterization step inherently differentiable; this leads to impeded rendering quality and relies on hyper-parameters to control the differentiability (i.e., support of non-zero gradient).}
The work related most closely to our approach in terms of gradient definition is the neural mesh renderer (NMR) \cite{kato2018neural}. 
We both construct $ \Phi_{\xx} $ depending on the change of pixel $ \Delta \Image_\xx $, but our method differs from NMR in the following aspects:
\begin{inparaenum}
\item we consider the movement of $ \pp_k $ in 3D space, while NMR only considers movement in the image plane, hence neglecting the gradient in $ z $-dimension. 
\item we define the gradient for all dimensions of $ \pp $ jointly. In contrast, NMR evaluates the 1D gradients separately and consequently considers only pixels in the same row and column;
\item we consider a set of occluded and occluding ellipses projected to pixel $ \xx $. This not only leads to more accurate gradient values, but also encourages noisy points inside the model to move onto the surface, to a position with matching pixel color.
\end{inparaenum}
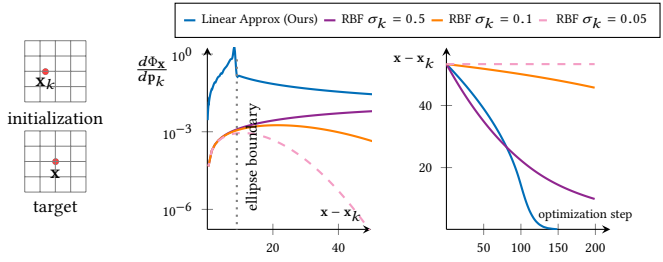
\begin{figure}
	\newlength{\imgheight}
	\setlength{\imgheight}{4cm}
	\begin{subfigure}{0.15\linewidth}
		\begin{tikzpicture}
		\footnotesize
		\node[anchor=south west,inner sep=0] (initial) {\includegraphics[height=0.2\imgheight, keepaspectratio, width=0.9\linewidth]{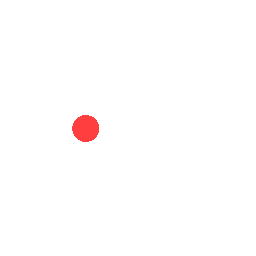}};
		\node[below=1pt of initial] (init cap) {initialization};
		\node[anchor=south west,inner sep=0,below=0pt of {init cap}] (target) {\includegraphics[width=0.9\linewidth, height=0.2\imgheight, keepaspectratio]{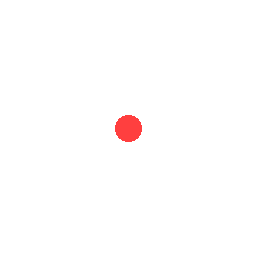}};
		\node[below=1pt of target] (target cap) {target};
		\begin{scope}[shift={($(initial.south west)$)},x={(initial.south east)},y={(initial.north west)}]
		\draw[help lines,xstep=.25,ystep=.25] (0,0) grid (1,1);
		\node[inner sep=0pt] at (0.35, 0.30) {$ \xx_{k} $};
		\end{scope}
		\begin{scope}[shift={($(target.south west)$)},x={(target.south east)},y={(target.north west)}]
		\draw[help lines,xstep=.25,ystep=.25] (0,0) grid (1,1);
		\node[inner sep=0pt] at (0.5, 0.30) {$ \xx $};
		\end{scope}
		\end{tikzpicture}
	\end{subfigure}\hspace{0.3cm}
	\begin{subfigure}{0.8\linewidth}
		\centering
		\begin{tikzpicture}
		\pgfplotsset{
			every tick label/.append style={font=\tiny},
every axis label/.append style={font=\tiny},
			width=0.55\linewidth, height=\imgheight, compat=1.3,
			legend image code/.code={
				\draw[mark repeat=1,mark phase=2]
				plot coordinates {
					(0cm,0cm)
					(0.1cm,0cm)        
					(0.15cm,0cm)         
				};%
			}
		}
		\begin{semilogyaxis}[
		axis x line = middle,
		axis y line = middle,
		ylabel style = {xshift=-8ex},
		name=plot1,
		xmin=0,xmax=50,
		legend style={at={(-0.2,1.15)},font=\tiny,anchor=west, legend columns=-1},
		xlabel={$ \xx-\xx_{k} $},
		ylabel={$\frac{d\Phi_{\xx}}{d\pp_{k}}$}
		]
		\addplot[mark=none, NavyBlue, thick, smooth] table [y index=0,header=false,x expr=128-\thisrowno{1}] {figures/pgf_plots/log.txt};\label{plt:ours}\addlegendentry{Linear Approx (Ours)}
		\addplot[mark=none, Plum, thick] table [y expr=\thisrowno{0},header=false,x expr=\thisrowno{1}] {figures/pgf_plots/real_grad_vrk0.5.txt};\label{plt:vrk0.5}\addlegendentry{RBF $ \sigma_k=0.5 $}
		\addplot[mark=none, orange, thick] table [y expr=\thisrowno{0},header=false,x expr=\thisrowno{1}] {figures/pgf_plots/real_grad_vrk0.1.txt};\label{plt:vrk0.1}\addlegendentry{RBF $ \sigma_k=0.1 $}
		\addplot[mark=none, dashed, Lavender, thick] table [y expr=\thisrowno{0},header=false,x expr=\thisrowno{1}] {figures/pgf_plots/real_grad_vrk0.05.txt};\label{plt:vrk0.05}\addlegendentry{RBF $ \sigma_k=0.05 $}
		\draw [thick, dotted, draw=gray] 
		(axis cs: 9,1e-8) -- (axis cs: 9,1) node[pos=0.2, anchor=north west, rotate=90] {\scriptsize ellipse boundary};
		\end{semilogyaxis}
		\begin{axis}[
		axis x line = middle,
		axis y line = middle,
		xlabel style = {xshift=3ex},
		ylabel style = {xshift=-6ex},
		at={($(plot1.east)+(1cm,0)$)},
		anchor=west,xmin=0, xmax=200,
		enlarge y limits={upper, value=0.1},
		enlarge x limits={upper, value=0.1},
		xlabel={optimization step},
		ylabel={$\xx-\xx_{k}$}
		]
		\addplot[mark=none, NavyBlue, thick, smooth] table [y expr=128-\thisrowno{1},header=false,x expr=\coordindex] {figures/pgf_plots/log.txt};\label{plt:ours_cvg}
		\addplot[mark=none, Plum, thick] table [y expr=128-\thisrowno{1},header=false,x expr=\coordindex] {figures/pgf_plots/one-point-vrk0.5.txt};\label{plt:vrk0.5_cvg}
		\addplot[mark=none, orange, thick] table [y expr=128-\thisrowno{1},header=false,x expr=\coordindex] {figures/pgf_plots/one-point-vkr0.1.txt};\label{plt:vrk0.1_cvg}
		\addplot[mark=none, dashed, Lavender, thick] table [y expr=128-\thisrowno{1},header=false,x expr=\coordindex] {figures/pgf_plots/one-point-vrk0.05.txt};\label{plt:vrk0.05_cvg}
		\end{axis}
		\end{tikzpicture}
	\end{subfigure}
	\caption{\chch Comparison between RBF-based gradient and our gradient approximation in terms of the gradient value at pixel $ \xx$ and residual in image space $ \xx-\xx_{k} $ as we optimize the point position $ \pp_k $ in the initial rendered image to match the target image. While our approximation (blue) is invariant under the choice of the hyper-parameter $ \sigma_k $, the RBF-based gradient (purple, orange and the dashed pink curves) is highly sensitive to its value. Small variations of $ \sigma_k $ can severely impact the convergence rate. }\label{fig:1Dplot}
\end{figure}
 \paragraph{Comparison to filter-based gradient approximation.} Alternatively, related to SoftRasterizer~\cite{liu2019soft} and Pix2Vex~\cite{petersen2019pix2vex}, one can define the gradient of the discontinuous function $ \Phi_{\xx, k} $ by replacing it with a $ C^\infty $ function, e.g., a radial basis function (RBF). 
This is a seemingly natural choice for EWA-based point rendering, since each point is represented as a RBF in the forward pass.
{\chch We compare the RBF-derived gradient with our approximation in a single point 1D scenario, and evaluate the gradient value and convergence rate.
As shown in \figref{fig:1Dplot}, the RBF-derived gradient is highly sensitive to the Gaussian filter's standard deviation $\sigma_k $. 
A small $ \sigma_k $ leads to diminishing gradient for distant pixels, causing convergence issues, as demonstrated with the dashed plot.
For a large $ \sigma_k $, $ \|\frac{d\Phi_{\xx}}{d\pp_{k}} \|$ can increase with $ \xx -\xx_{k} $ when the pixel is outside the ellipse boundary; as a result, the optimization is prone to fall into a local minima in multi-point scenario as shown in \figref{fig:twopoints}.
Lastly, it is not obvious how to extend the RBF derivation for the depth dimension, while the linear approximation naturally applies to all dimensions.}

\subsection{Surface regularization}
The lack of structure in point clouds, while providing freedom of massive topology changes, can pose a significant challenge for optimization.
First, the gradient derivation is entirely paralleled; as a result, points move irrespective of each other.
Secondly, as the movement of points will only induce small and sparse changes in the rendered image, gradients on each point are less structured compared to corresponding gradients for meshes.
Without proper regularization, one can quickly end up in local minima.

Inspired by \cite{Huang2009wlop,oztireli2009feature}, we propose regularization to address this problem based on two parts: a repulsion and a projection term.
The repulsion term is aimed at generating uniform point distributions by maximizing the distances between its neighbors on a local projection plane, while the projection term preserves clean surfaces by minimizing the distance from the point to the surface tangent plane.

Obviously, both terms require finding a reliable surface tangent plane.
However, this can be challenging, since during optimization, especially in the case of multi-view joint optimization, intermediate point clouds can be very noisy and contain many occluded points inside the model, hence we propose a weighted PCA to penalize the occluded inner points.
In addition to the commonly used bilateral weights which considers both the point-to-point euclidean distance and the normal similarity, we propose a visibility weight, which penalizes occluded points, since they are more likely to be outliers inside the model.

Let $ \pp_{i} $ denote a point in question and $ \pp_{k} $ denote one point in its neighborhood, $ \pp_{k}\in\lbrace\pp | \  \|\pp-\pp_{i}\| \leq \DD\rbrace $, we propose computing a weighted PCA using the following weights
\begin{align}
 \psi_{ik} & = \exp\left(-\frac{\|\pp_i-\pp_{k}\|^2}{\DD^2}\right)\label{eq:psi}\\
 \theta_{ik} & = \exp\left(-\frac{\left(1-\nn_{k}^\tr\nn_i\right)^2}{\max\left(1e^{-5}, 1-\cos\left(\Theta\right)\right)}\right) \label{eq:theta}\\
\phi_{ik} & = \frac{1}{o_k+1},\label{eq:phi}
\end{align}
where $ \psi_{ik} $ and $ \theta_{ik} $ are bilateral weights which favor neighboring points that are spatially close and have similar normal orientation respectively, and $ \phi_{ik} $ is the proposed visibility weight which is defined using an occlusion counter $ o_k $ that counts the number of times $ \pp_{k} $ is occluded in all camera views.
Then a reliable projection plane can be obtained using singular value decomposition from weighted vectors $ w_{ik} \left(\pp_{i} - \sum_{k=0}^{K}w_{ik} \pp_{k}\right)$, where $ w_{ik} = \frac{\psi_{ik}\theta_{ik}\phi_{ik}}{\sum_{i=0}^{K} \psi_{ik}\theta_{ik}\phi_{ik}}$.

For the repulsion term, the projected point-to-point distance is obtained via $ d_{ik} = \tilde{\VV}\tilde{\VV}^\tr\left(\pp_i - \pp_{k}\right) $, where $ \tilde{\VV} $ contains the first $ 2 $ principal components.
We define the repulsion loss as follows and minimize it together with the per-pixel image loss
\begin{equation}
\LL_{r} = \frac{1}{N}\sum_{N}\sum_{K}\frac{\psi_{ik}}{d_{ik}^2+10^{-4}}.
\end{equation} 

For the projection term, we minimize the point-to-plane distance via $ d_{ik} = \VV_n\VV^\tr\left(\pp_{i}-\pp_{k}\right)$, where $ \VV_n $ is the last components.
Correspondingly, the projection loss is defined as 
\begin{equation}\label{eq:projLoss}
\LL_{p} = \frac{1}{N}\sum_{N}\sum_{K}w_{ik}d_{ik}^2.
\end{equation}
\begin{figure}
	\setlength{\tabcolsep}{1pt}
	\begin{tabular}{cccc}
		\makecell[t]{initialization\newline} & \makecell{target\newline} & \makecell{without\\repulsion} & \makecell{with\\repulsion} \\
		\begin{subfigure}[t]{0.24\linewidth}
			\includegraphics[width=\linewidth,clip,trim={1cm, 4cm, 1cm, 4cm}]{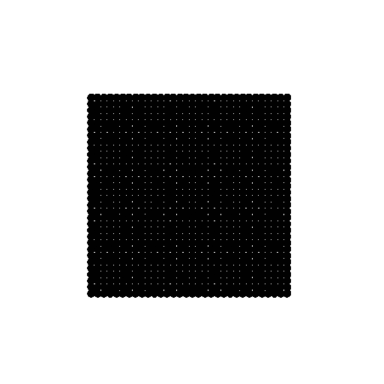}
		\end{subfigure}&
		\begin{subfigure}[t]{0.26\linewidth}
			\includegraphics[width=\linewidth,clip,trim={1.5cm, 4cm, 1.0cm, 5cm}]{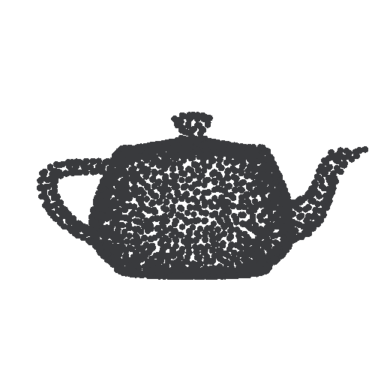}
		\end{subfigure}&%
		\begin{subfigure}[t]{0.2\linewidth}
			\includegraphics[width=\linewidth,clip,trim={1.5cm, 4cm, 1.cm, 5cm}]{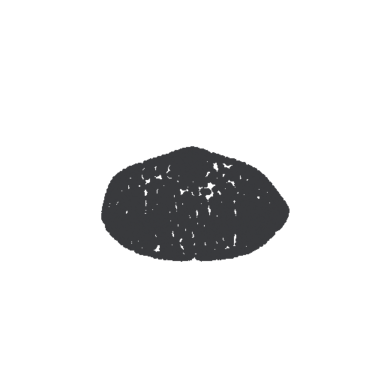}
		\end{subfigure}&%
		\begin{subfigure}[t]{0.26\linewidth}
			\includegraphics[width=\linewidth,clip,trim={1.5cm, 4cm, 1.cm, 5cm}]{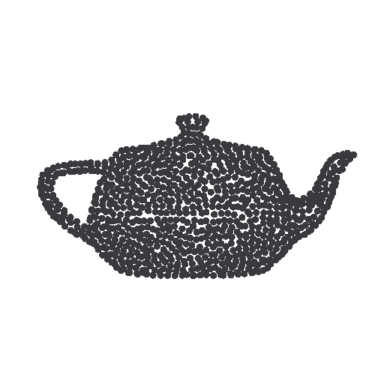}
		\end{subfigure}
	\end{tabular}
	\caption{The effect of repulsion regularization. We deform a 2D grid to the teapot. Without the repulsion term, points cluster in the center of the target shape. The repulsion term penalizes this type of local minima and encourages a uniform point distribution.}\label{fig:2Dlocalminima}
\end{figure}

\begin{figure}
	\setlength{\tabcolsep}{0pt}
	\begin{tabular}{cccc}
		\multicolumn{2}{c}{without projection term} &  \multicolumn{2}{c}{with projection term} \\
		\begin{subfigure}[t]{0.25\linewidth}
			\includegraphics[width=\linewidth]{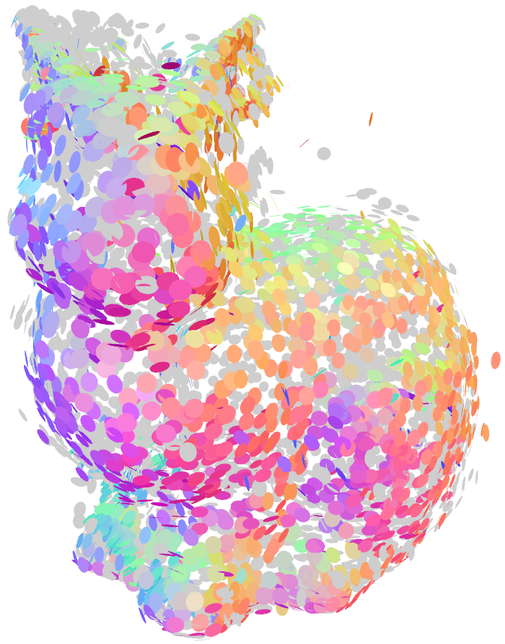}
		\end{subfigure}&
		\begin{subfigure}[t]{0.25\linewidth}
			\includegraphics[width=\linewidth]{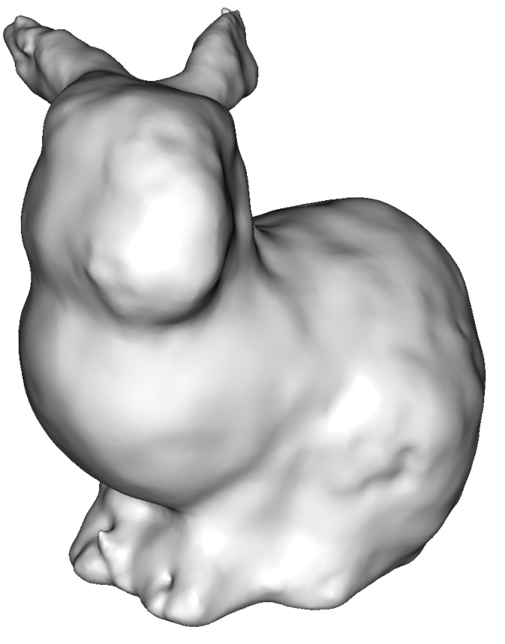}
		\end{subfigure}&
		\begin{subfigure}[t]{0.25\linewidth}
			\includegraphics[width=\linewidth]{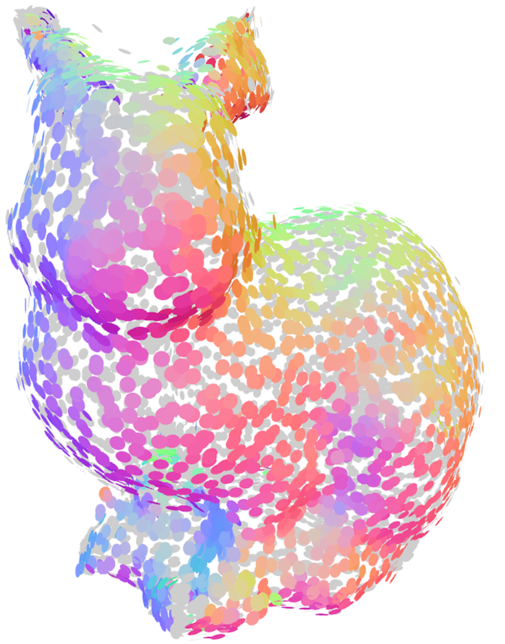}
		\end{subfigure}&
		\begin{subfigure}[t]{0.25\linewidth}
			\includegraphics[width=\linewidth]{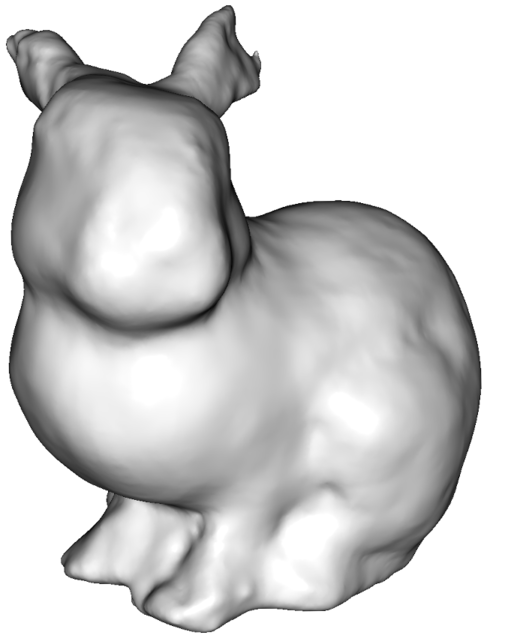}
		\end{subfigure}
	\end{tabular}
	\caption{The effect of projection regularization. The projection term effectively enforces points to form a local manifold. For a better visualization of outliers inside and outside of the object, we use a small disk radius and render the backside of the disks using light gray color.}\label{fig:projection}
\end{figure}
The effect of repulsion and projection terms are clearly demonstrated in \figref{fig:2Dlocalminima} and \figref{fig:projection}.
In \figref{fig:2Dlocalminima}, we aim to move points lying on a 2D grid to match the silhouette of a 3D teapot. 
Without the repulsion term, points quickly shrink to the center of the reference shape, which is a common local minima since the gradient coming from surrounding pixels cancel each other out.
With the repulsion term, the points can escape such local minima and distribute evenly inside the silhouette.
In \figref{fig:projection} we deform a sphere to bunny from 12 views. 
Without projection regularization, points are scattered within and outside the surface.
In contrast, when the projection term is applied, we can obtain a clean and smooth surface.
\section{Implementation details.}\label{sec:implementation}
\subsection{Optimization objective}
We choose Symmetric Mean Absolute Percentage Error (SMAPE) as the image loss $ \LL_\Image $. 
SMAPE is designed for high dynamic range images such as rendered images therefore it behaves more stable for unbounded values \cite{vogels2018denoising}.
It is defined as 
\begin{equation}\label{eq:smape}
\LL_\Image = \dfrac{1}{HW}\sum_{\xx\in \Image}\sum_{c}^{C}\dfrac{|\Image_{\xx,c}-\Image^*_{\xx,c}|}{|\Image_{\xx,c}|+|\Image^*_{\xx,c}|+\epsilon}, 
\end{equation}
where $ H $ and $ W $ are the dimensions of the image, the value of $ \epsilon $ is typically chosen as $ 10^{-5} $.

The total optimization objective corresponding to~\equref{eq:backward_overview} for a set of views $ V $ amounts to
\begin{equation}\label{eq:total_loss}
\sum_{v=0}^{V}\LL\left(\Image_v,\Image_v^*\right) = \sum_{v=0}^{V}\LL_\Image\left(\Image_v,\Image_v^*\right)+\gamma_p\LL_{p}+\gamma_r\LL_{r}.
\end{equation}
Loss weights $ \gamma_p $ and $ \gamma_r $ are typically chosen to be $ 0.02, 0.05 $ respectively.

\subsection{Alternating normal and point update}
For meshes, the face normals are determined by point positions.
For points, though, normals and point positions can be treated as independent entities thus optimized individually.
Our pixel value factorization in~\equref{eq:gradP} and \equref{eq:gradN} means that, the gradient on point positions $ \pp $ mainly stems from the visibility term, while gradients on normals $ \nn $ can be derived from $ \ww_k $ and $ \rho_{k} $. 
Because the gradient w.r.t.\ $ \nn $ and $ \pp $ assumes the other stays fixed, we apply the update of $ \nn $ and $ \pp $ in an alternating fashion.
Specifically, we start with normals, execute optimization for $ T_\nn $ times then we optimize point positions for $ T_\pp $ times. 

As observed in many point denoising works \cite{oztireli2009feature,Huang2009wlop,guerrero2018pcpnet}, finding the right normal is the key for obtaining clean surfaces.
Hence we efficiently utilize the improved normals even if the point positions are not being updated, in that we directly update the point positions using the gradient from the regularization terms $ \frac{\partial \LL_{p}}{\partial \pp_{k}} $ and $ \frac{\partial \LL_{r}}{\partial \pp_{k}} $.
In fact, for local shape surface modification, this simple strategy consistently yields satisfying results.

\subsection{Error-aware view sampling}\label{sec:view_sampling}
View selection is very important for quick convergence.
In our experiments, we aim to cover all possible angles by sampling camera positions from a hulling sphere using farthest point sampling.
Then we randomly perturb the sampled position and set the camera to look at the center of the object.
The sampling process is repeated periodically to further improve optimization.

However, for shapes with complex topology, such a sampling scheme is not enough.
We propose an error-aware view sampling scheme which chooses the new camera positions based on the current image loss.

Specifically, we downsample the reference image and the rendered result, then compute the pixel position with the largest image error.
Then we find $ K $ points whose projection is closest to the found pixel. 
The mean 3D position of these points will be the center of focus.
Finally, we sample camera positions on a sphere around this focal point with a relatively small distance.
Such techniques help us to improve point positions in small holes during large shape deformation.

\section{Results}
\label{sec:results}
\begin{figure}\centering
	\begin{subfigure}{\linewidth}
		\setlength{\tabcolsep}{0pt}
		\begin{tabular}{ccccc}
			&initialization & target & result & Meshlab render\\
			\rotatebox[origin=l]{90}{\hspace{3ex}Paparazzi}& 
			\includegraphics[width=0.20\linewidth,trim={0cm, 0cm, 0cm, 1.5cm},clip]{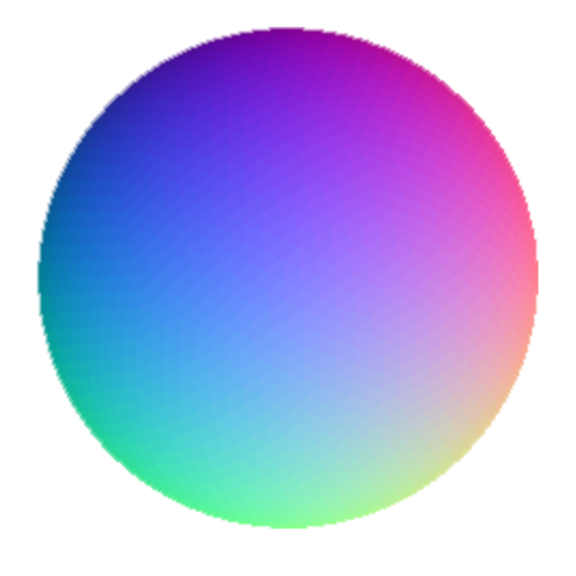}&
			\includegraphics[width=0.24\linewidth,trim={0, 0cm, 0cm, 1.5cm},clip]{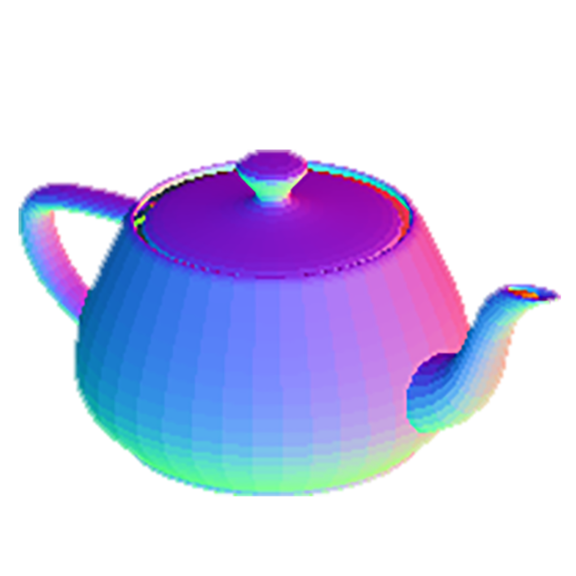}&
			\includegraphics[width=0.20\linewidth,trim={0, 0cm, 0cm, 1.5cm},clip]{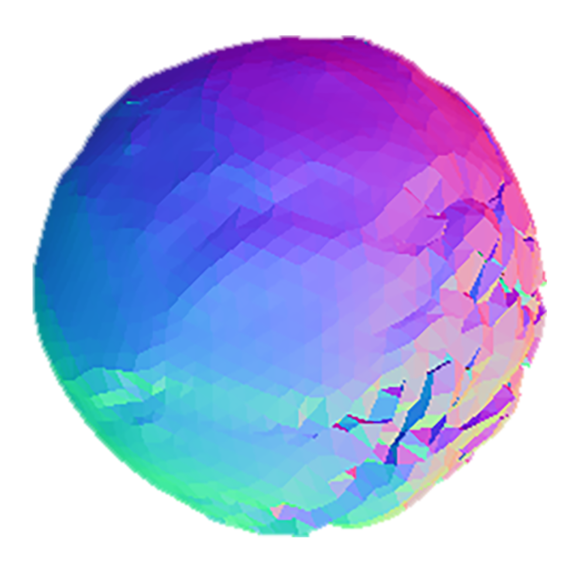}&
			\includegraphics[width=0.20\linewidth,trim={0, 0cm, 0cm, 1.5cm},clip]{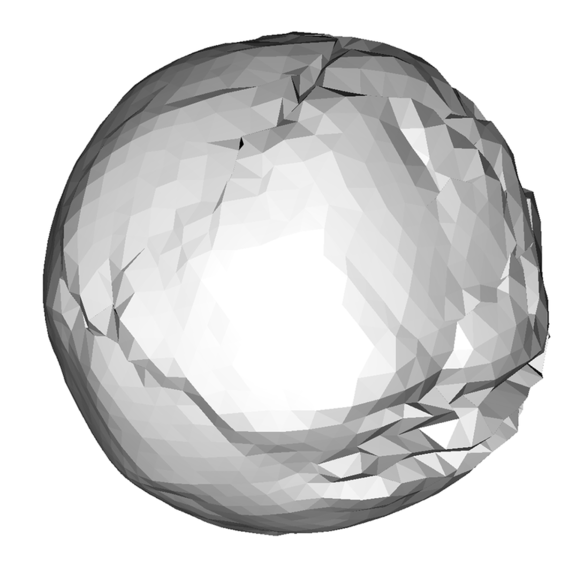}\\      
			\rotatebox[origin=l]{90}{\hspace{4ex}NMR}&
			\includegraphics[width=0.20\linewidth,trim={0, 1.5cm, 0cm, 1.5cm},clip]{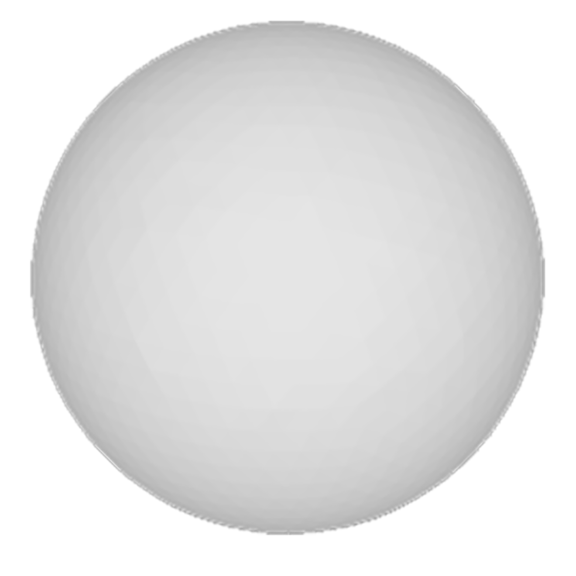}&
			\includegraphics[width=0.24\linewidth,trim={0, 2.5cm, 0cm, 2.5cm},clip]{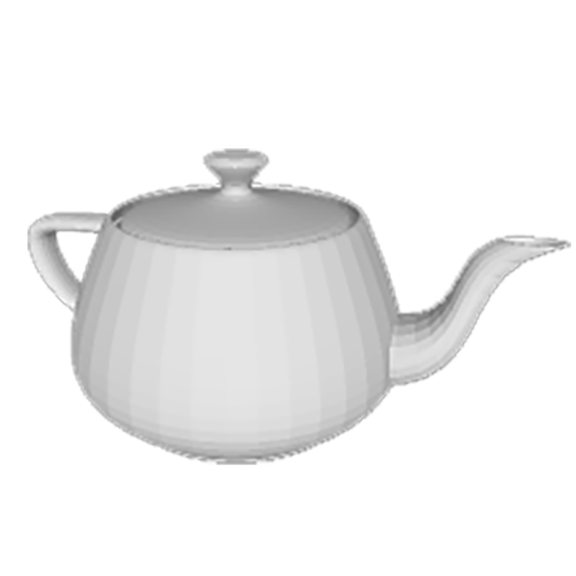}&
			\includegraphics[width=0.24\linewidth,trim={0, 1.5cm, 0cm, 1.5cm},clip]{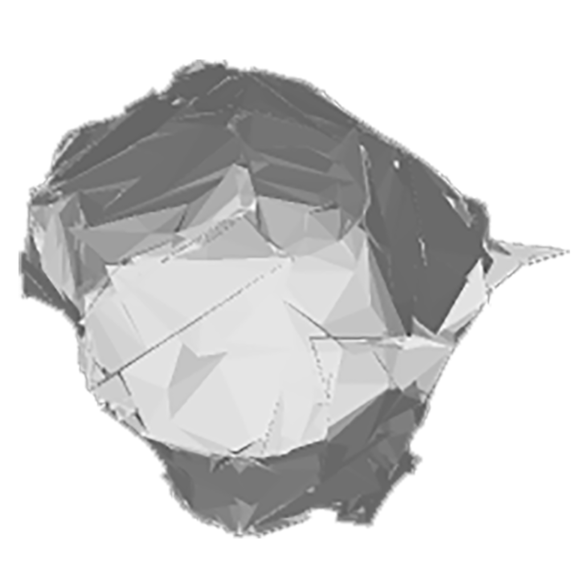}&
			\includegraphics[width=0.24\linewidth,trim={0, 1.5cm, 0cm, 1.5cm},clip]{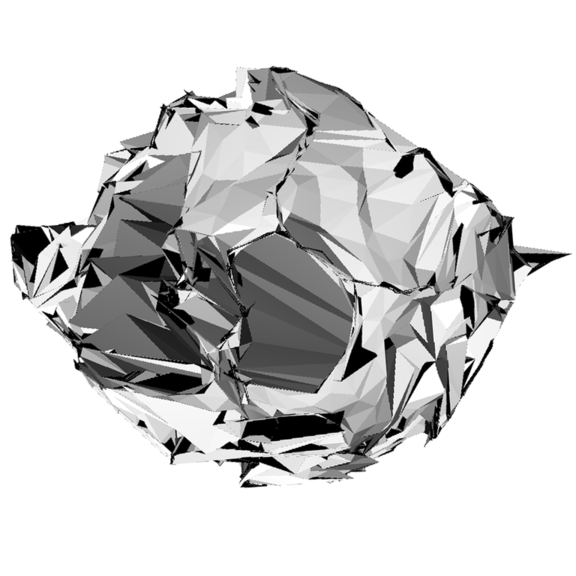}\\
			\rotatebox[origin=l]{90}{\hspace{2ex}OpenDR}&
			\includegraphics[width=0.22\linewidth,trim={.5cm, .5cm, .5cm, .5cm},clip]{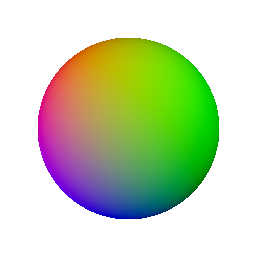}&
			\includegraphics[width=0.24\linewidth,trim={0, 1cm, 0.5cm, 2cm},clip]{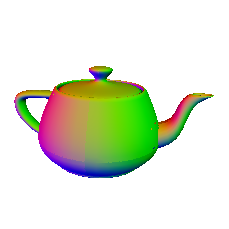}&
			\includegraphics[width=0.24\linewidth,trim={0, 1cm, 0cm, 2cm},clip]{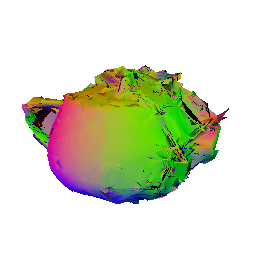}&
			\includegraphics[width=0.24\linewidth,trim={0, 7cm, 0cm, 3.5cm},clip]{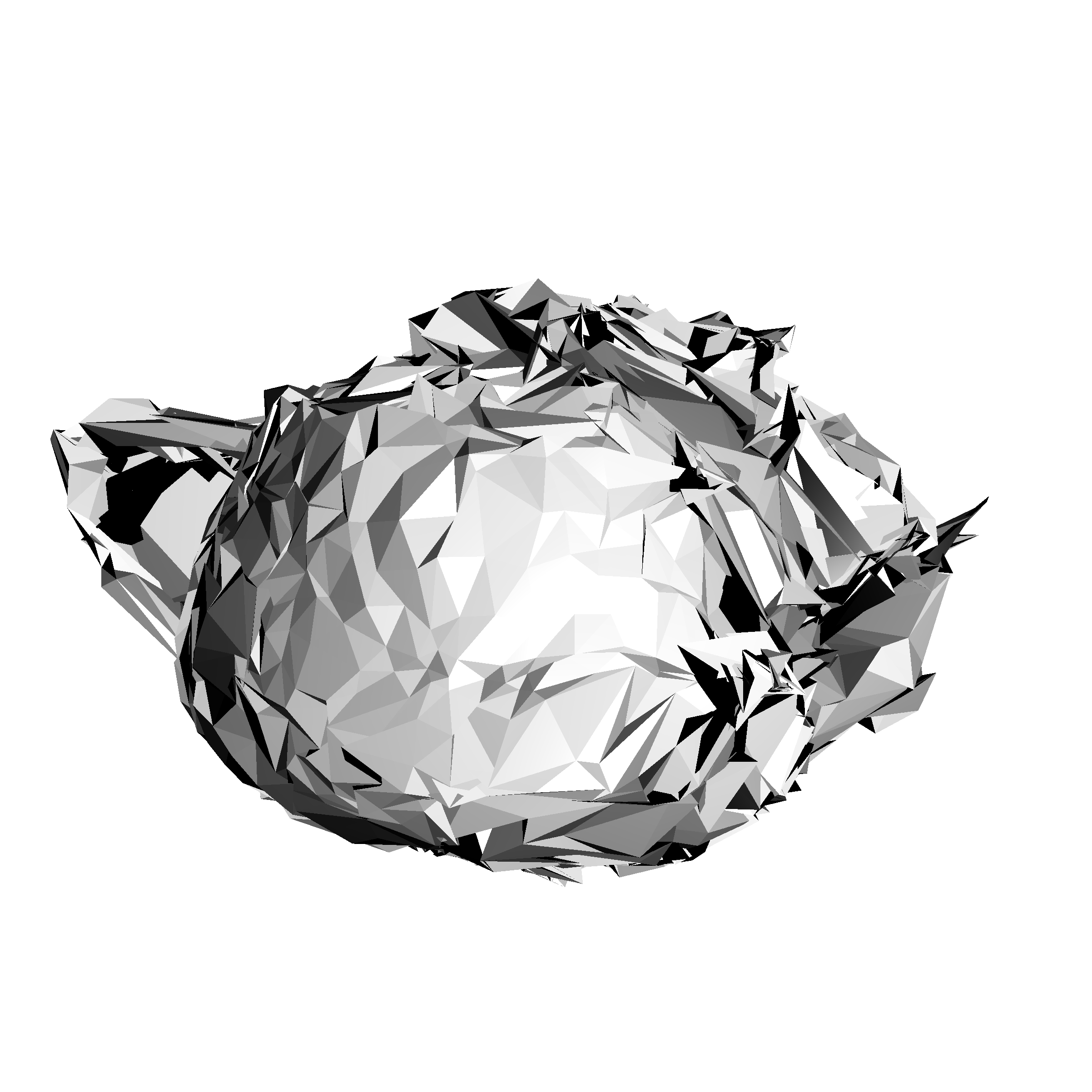}\\   
			\rotatebox[origin=l]{90}{\hspace{4ex}Ours}&
			\includegraphics[width=0.20\linewidth,trim={0, 0cm, 0cm, 1.5cm},clip]{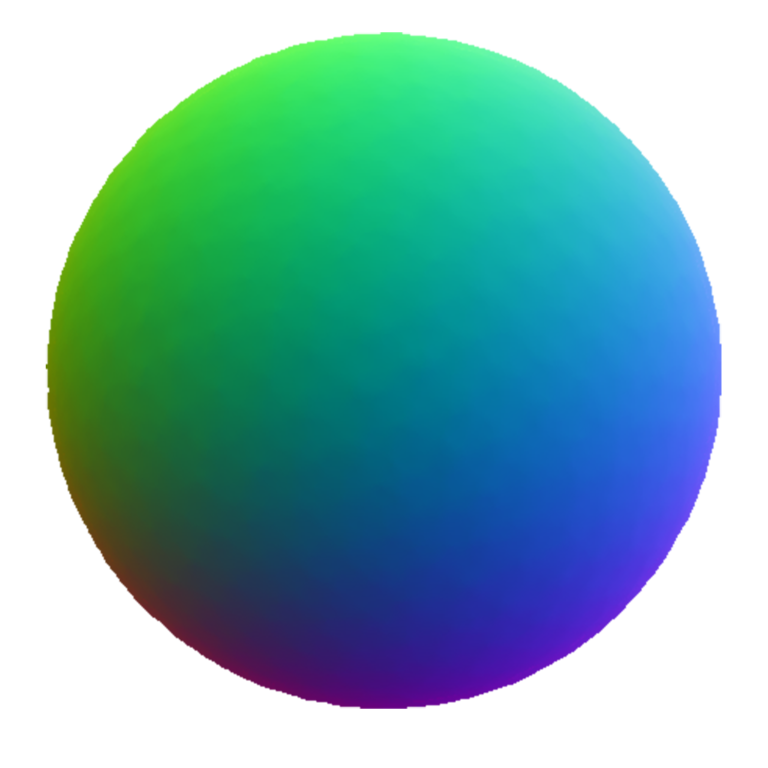}&
			\includegraphics[width=0.24\linewidth,trim={0, 1.5cm, 0cm, 1.5cm},clip] {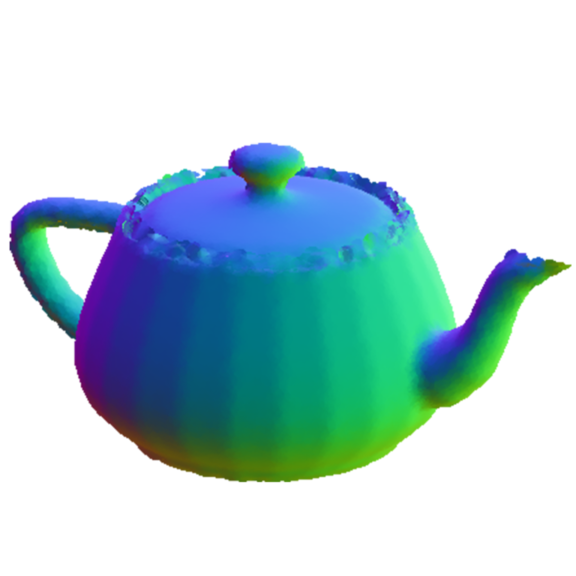}&
			\includegraphics[width=0.24\linewidth,trim={0, 1.5cm, 0cm, 1.5cm},clip] {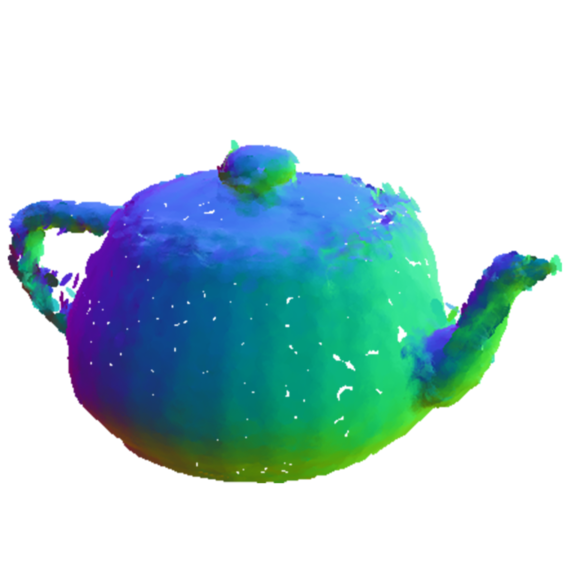}&
			\includegraphics[width=0.24\linewidth,trim={0, 1.5cm, 0cm, 1.5cm},clip] {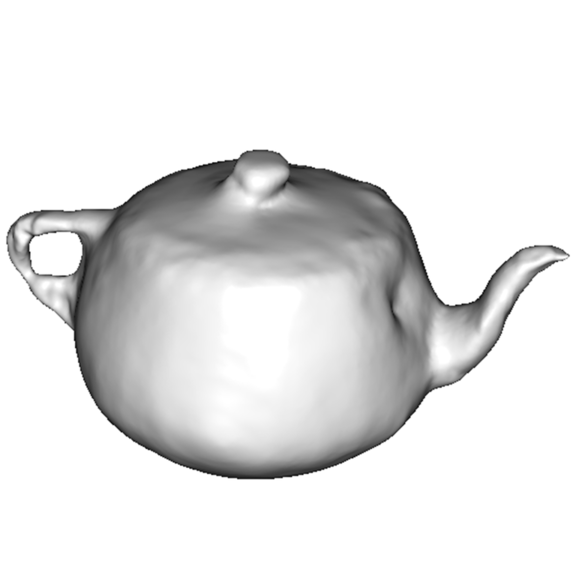}
		\end{tabular}
	\end{subfigure}
	\caption{\chch Large shape deformation with topological changes, compared with three mesh-based DRs, namely Paparazzi \cite{liu2018paparazzi}, OpenDR \cite{loper2014opendr} and Neural Mesh Renderer \cite{kato2018neural}. Compared to the mesh-based approaches, DSS faithfully recovers the handle and cover of the teapot thanks to the flexibility of the point-based representation.}
	\label{fig:compare_topology}
\end{figure}
We evaluate the performance of DSS by comparing it to state-of-the-art DRs, and demonstrate its applications in point-based geometry editing and filtering. 

Our method is implemented in Pytorch~\cite{paszke2017automatic}, we use stochastic gradient descent with Nesterov momentum \cite{sutskever2013importance} for optimization. 
A learning rate of $ 5 $ and $ 5000 $ is used for points and normals, respectively. 
In all experiments, we render in back-face culling mode with $ 256 \times 256 $ resolution and diffuse shading, using RGB sun lights fixed relative to the camera position. 

Unless otherwise stated, we optimize for up to 16 cycles of $ T_\nn $ and $ T_\pp $ optimization steps for point normal and position (for large deformation $ T_\pp = 25 $ and $ T_\nn = 15 $; for local surface editing $ T_\nn = 19 $ and $ T_\pp = 1 $).
In each cycle, 12 randomly sampled views are used simultaneously for an optimization step.
To test our algorithms for noise resilience, we use random white Gaussian noise with a standard deviation measured relative to the diagonal length of the bounding box of the input model. We refer to Appendix~\ref{sec:parameters} for a detailed discussion of parameter settings.

\subsection{Comparison of different DRs.}\label{sec:comparison}
\begin{figure}
	\centering
	\includegraphics[width=\linewidth]{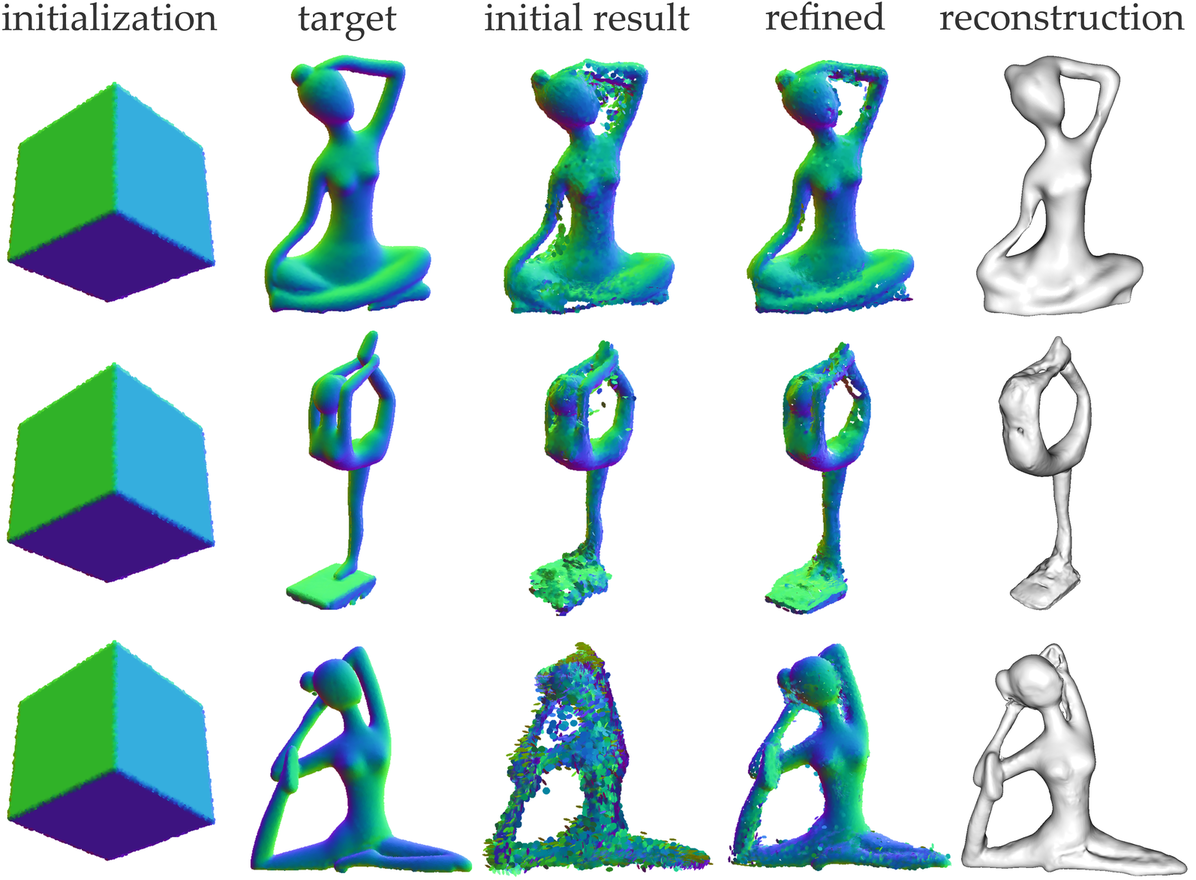}
	\caption{DSS deforms a cube to three different Yoga models. Noisy points may occur when camera views are under-sampled or occluded (as shown in the initial result). We apply an additional refinement step improving the view sampling as described in \secref{sec:view_sampling}.}
	\label{fig:yoga}
\end{figure}
\begin{figure}
\centering
\includegraphics[width=\linewidth]{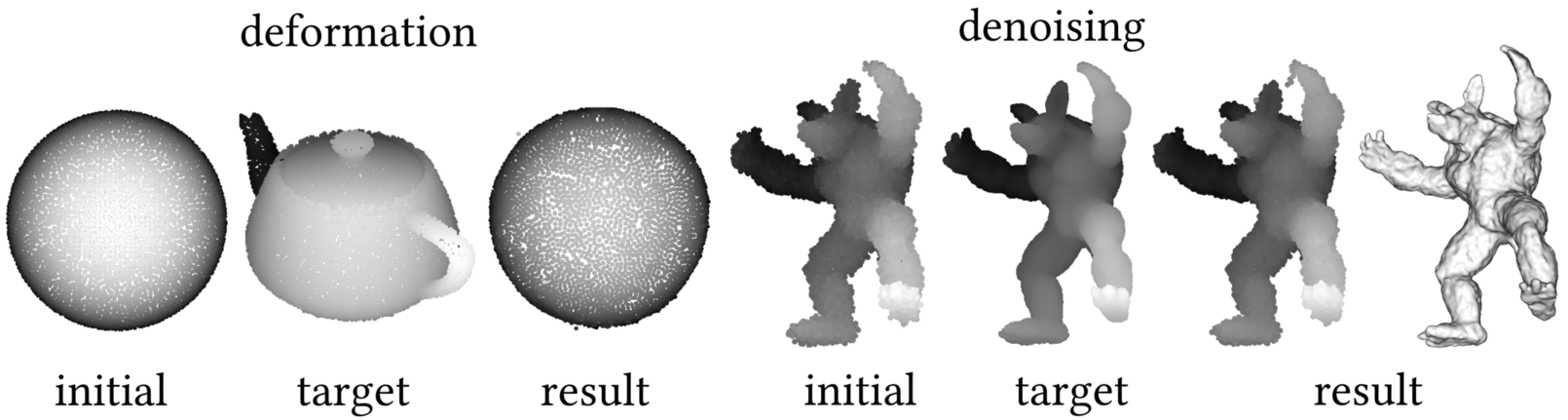}
\caption{A simple projection-based point renderer which renders depth values fails in deformation and denoising tasks.}
\label{fig:pseudo}
\end{figure}
We compare DSS in terms of large geometry deformation to the state-of-the-art mesh-based DRs, i.e., OpenDR~\cite{loper2014opendr}, NMR~\cite{kato2018neural} and Paparazzi~\cite{liu2018paparazzi}.
For the mesh DRs, we use the publicly available code provided by the authors and report the best results among experiments using different parameters (e.g., number of cameras and learning rate). 
{\changed 
All methods use the same initial and target shape, and similar camera positions. 

Among the mesh-based methods, OpenDR can best deform an input sphere to match the silhouette of a target teapot. However, none of these methods can handle topology changes (see the handle) and struggle with large deformation (see the spout).
In comparison, DSS recovers these geometry structures with high fidelity and at the same time produces more elaborate surface details (see the pattern on the body of the teapot).}

Finally, we compare with a naive point DR based on \cite{roveri2018pointpronets,roveri2018network,insafutdinov2018unsupervised}, where the pixel intensities are represented by the sum of smoothed depth values.
As shown in \figref{fig:pseudo}, such a naive implementation of point-based DR cannot handle large-scale shape deformation nor fine-scale denoising, because position gradient is confined locally restricting long-range movement and normal information is not utilized to fine-grained geometry update.

\subsection{Application: shape editing via image filter}

\begin{figure}
	\centering
	\includegraphics[width=\linewidth]{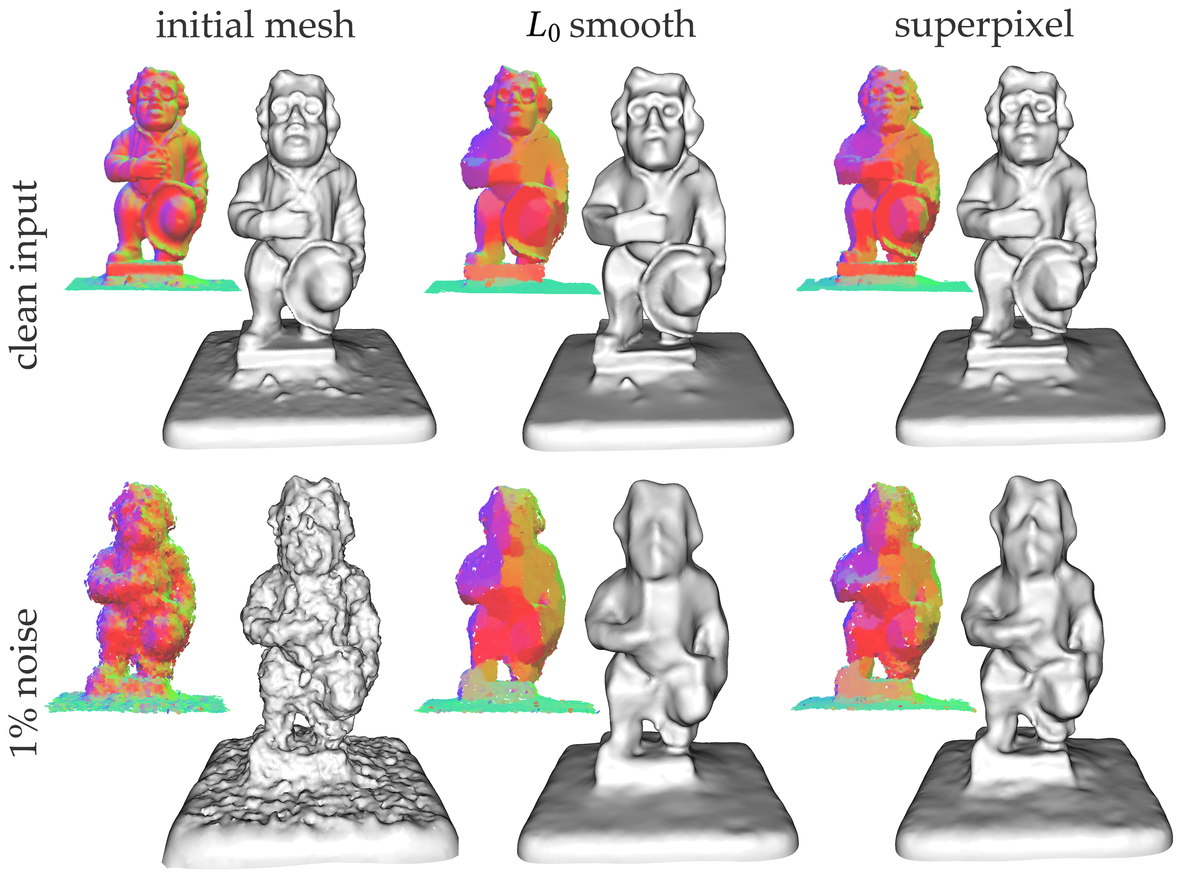}
	\caption{Examples of DSS-based geometry filtering. We apply image filters on the DSS rendered multi-view images and propagate the changes of pixel values to point positions and normals. From left to right are the Poisson reconstruction of input points, points filtered by $L_0$-smoothing, and superpixel segmentation. In the first row, a clean point cloud is used as input, while in the second row, we add 1\% white Gaussian noise. In both cases, DSS can update the geometry accordingly to match the changes in the image domain.}\label{fig:image_filtering}
\end{figure}

\begin{figure}
	\setlength{\tabcolsep}{1pt}
	\small
	\begin{tabular}{cccc}
		\makecell{clean\\initialization} & \makecell{$L_0$ smooth} & \makecell{noisy\\initialization} & \makecell{$L_0$ smooth}\\
		\begin{subfigure}{0.24\linewidth}
			\includegraphics[width=\linewidth,trim={1cm, 5cm, 1cm, 3cm},clip]{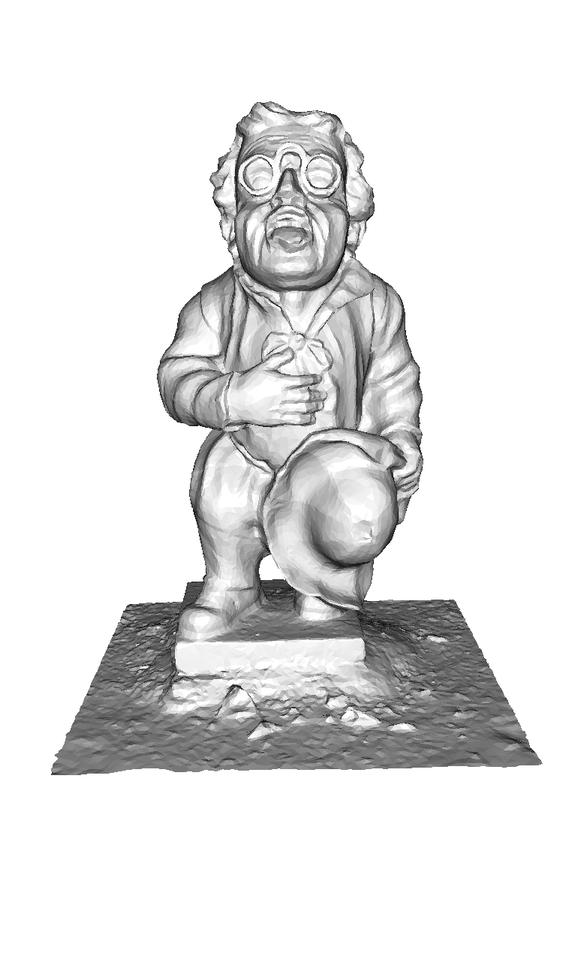}
		\end{subfigure}&
		\begin{subfigure}{0.24\linewidth}
			\includegraphics[width=\linewidth,trim={1cm, 5cm, 1cm, 3cm},clip]{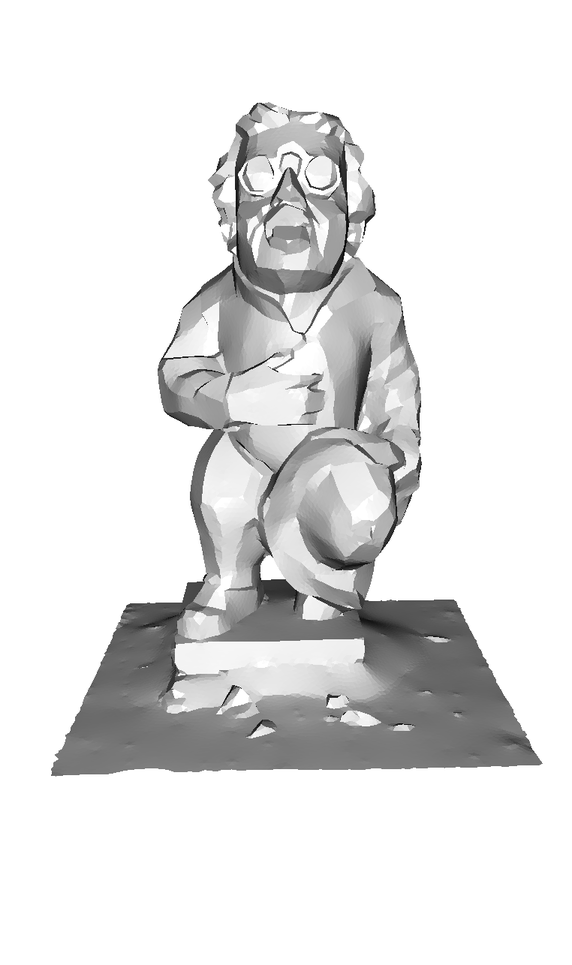}
		\end{subfigure}&
		\begin{subfigure}{0.24\linewidth}
			\includegraphics[width=\linewidth,trim={1cm, 5cm, 1cm, 3cm},clip]{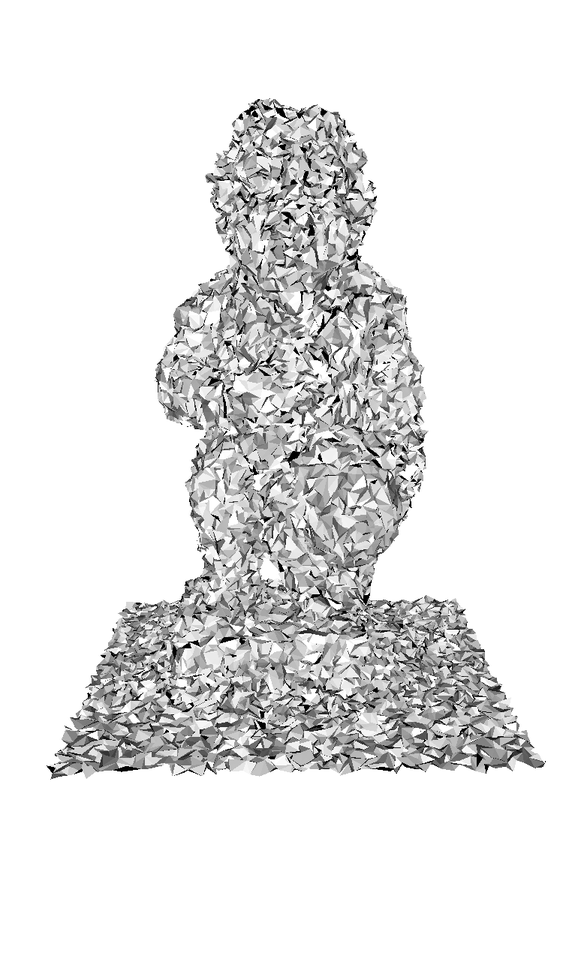}
		\end{subfigure}&
		\begin{subfigure}{0.24\linewidth}
			\includegraphics[width=\linewidth,trim={1cm, 5cm, 1cm, 3cm},clip]{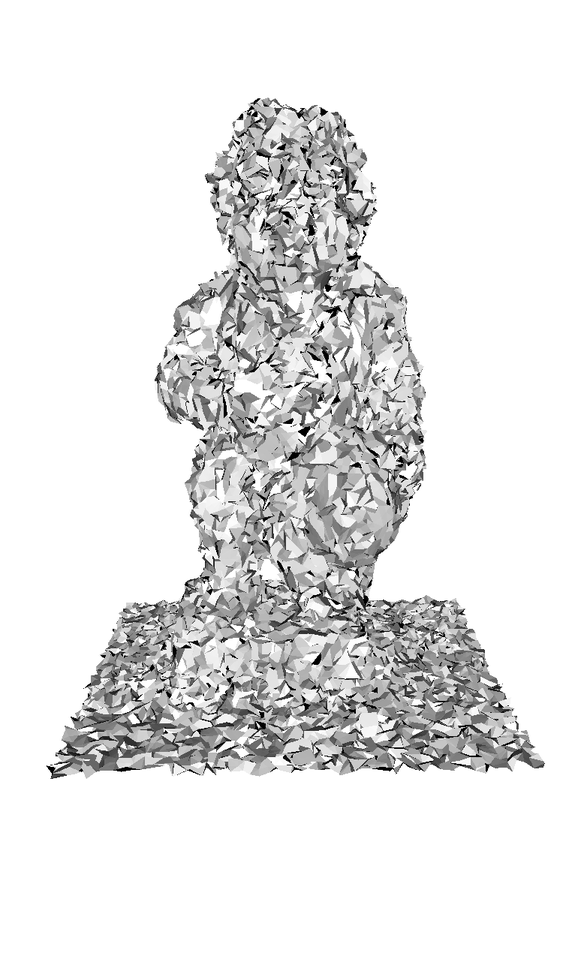}
		\end{subfigure}
	\end{tabular}
	\caption{Paparazzi~\cite{liu2018paparazzi} successfully applies a $L_0$ image filter to a clean mesh (Left) but fails on an input containing 0.5 \% noise (Right).}\label{fig:papa_filter}
\end{figure}
As demonstrated in Paparazzi, one important application of DR is shape editing using existing image filters. It allows many kinds of geometric filtering and style transfer, which would have been challenging to define purely in the geometry domain. 
This benefit also applies to DSS. 

We experimented with two types of image filters, L0 smoothing~\cite{xu2011image} and superpixel segmentation~\cite{achanta2012slic}. 
These filters are applied to the original rendered images to create references.
Like Paparazzi, we keep the silhouette of the shape and change the local surface geometry by updating point normals, then the projection and repulsion regularization are applied to correct the point positions.

As shown in \figref{fig:image_filtering}, DSS successfully transfers image-level changes to geometry.
Even under 1\% noise, DSS continues to produce reasonable results.
In contrast, mesh-based DRs are sensitive to input noise, because it leads to small piecewise structures and flipped faces in image space (see \figref{fig:papa_filter}), which are troublesome for the computation of gradients. 
In comparison, points are free of any structural constraints; thus, DSS can update normals and positions independently, which makes it robust under noise.

\begin{figure*}[h!]
	\setlength{\tabcolsep}{0pt}
	\begin{tabular}{ccccccccc}
		&\multicolumn{4}{c}{0.3\% noise}&\multicolumn{4}{c}{1.0\% noise}\\
		\rotatebox[origin=l]{90}{\hspace{4ex}input}&
		\includegraphics[width=0.11\linewidth]{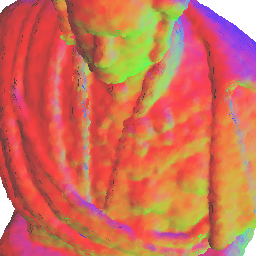}&
		\includegraphics[width=0.11\linewidth]{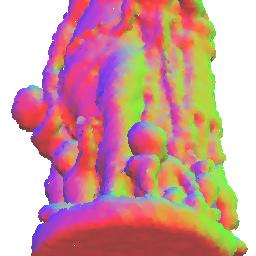}&
		\includegraphics[width=0.11\linewidth]{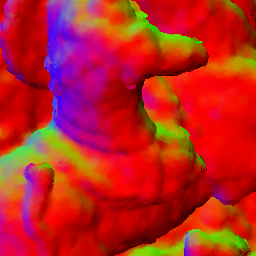}&
		\includegraphics[width=0.11\linewidth]{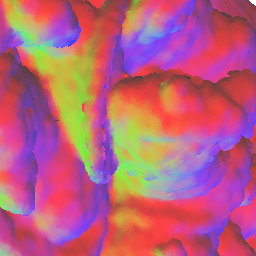}&
		
		\includegraphics[width=0.11\linewidth]{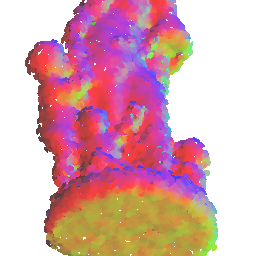}&
		\includegraphics[width=0.11\linewidth]{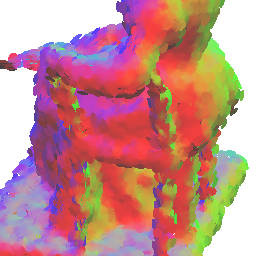}&
		\includegraphics[width=0.11\linewidth]{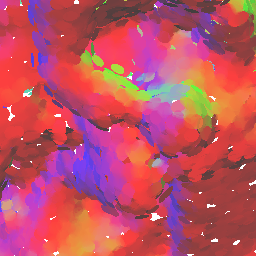}&
		\includegraphics[width=0.11\linewidth]{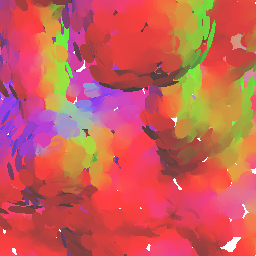}
		\\
		\rotatebox[origin=l]{90}{\hspace{3ex}output}&
		\includegraphics[width=0.11\linewidth]{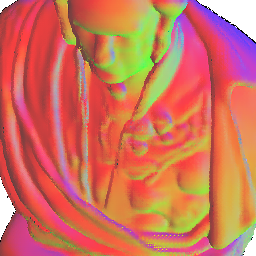}&
		\includegraphics[width=0.11\linewidth]{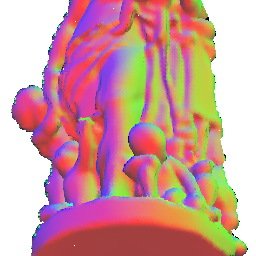}&
		\includegraphics[width=0.11\linewidth]{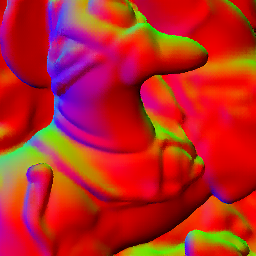}&
		\includegraphics[width=0.11\linewidth]{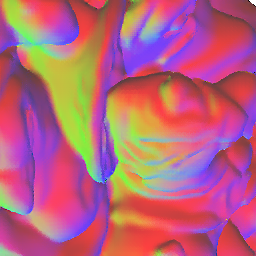}&
		
		\includegraphics[width=0.11\linewidth]{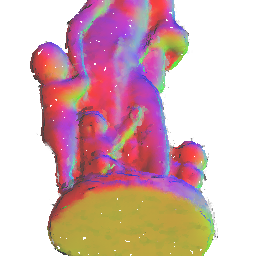}&
		\includegraphics[width=0.11\linewidth]{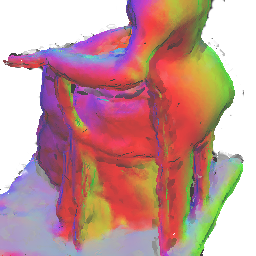}&
		\includegraphics[width=0.11\linewidth]{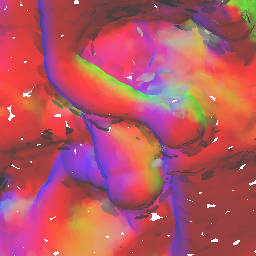}&
		\includegraphics[width=0.11\linewidth]{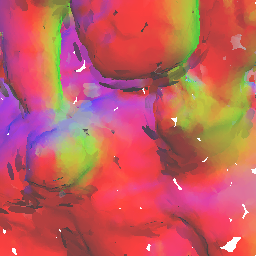}
	\end{tabular}
	\caption{Examples of the input and output of the Pix2Pix denoising network. We train two models to target two different noise levels (0.3\% and 1.0\%). In both cases, the network is able to recover smoothly detailed geometry, while the 0.3\% noise variant generates more fine-grained details. }\label{fig:pix2pix_output}
\end{figure*}
\subsection{Application: point cloud denoising}
One of the benefits of the shape-from-rendering framework is the possibility to leverage powerful neural networks and vast 2D data.
We demonstrate this advantage in a point cloud denoising task, which is known to be an ill-posed problem where handcrafted priors struggle with recovering all levels of smooth and sharp features. 

First, we train an image denoising network based on the Pix2Pix~\cite{isola2017image} framework, which utilizes the generative adversarial network \cite{goodfellow2014generative} to add plausible details for improved visual quality (we refer readers to Appendix for further details on the training data preparation as well as the adapted network architecture).
During test time, we render images of the noisy point cloud from different views and use the trained Pix2Pix network to reconstruct geometric structure from the noisy images.
Finally, we update the point cloud using DSS with the denoised images as reference.

\begin{figure*}[h!]
	\centering
	\includegraphics[width=0.9\linewidth]{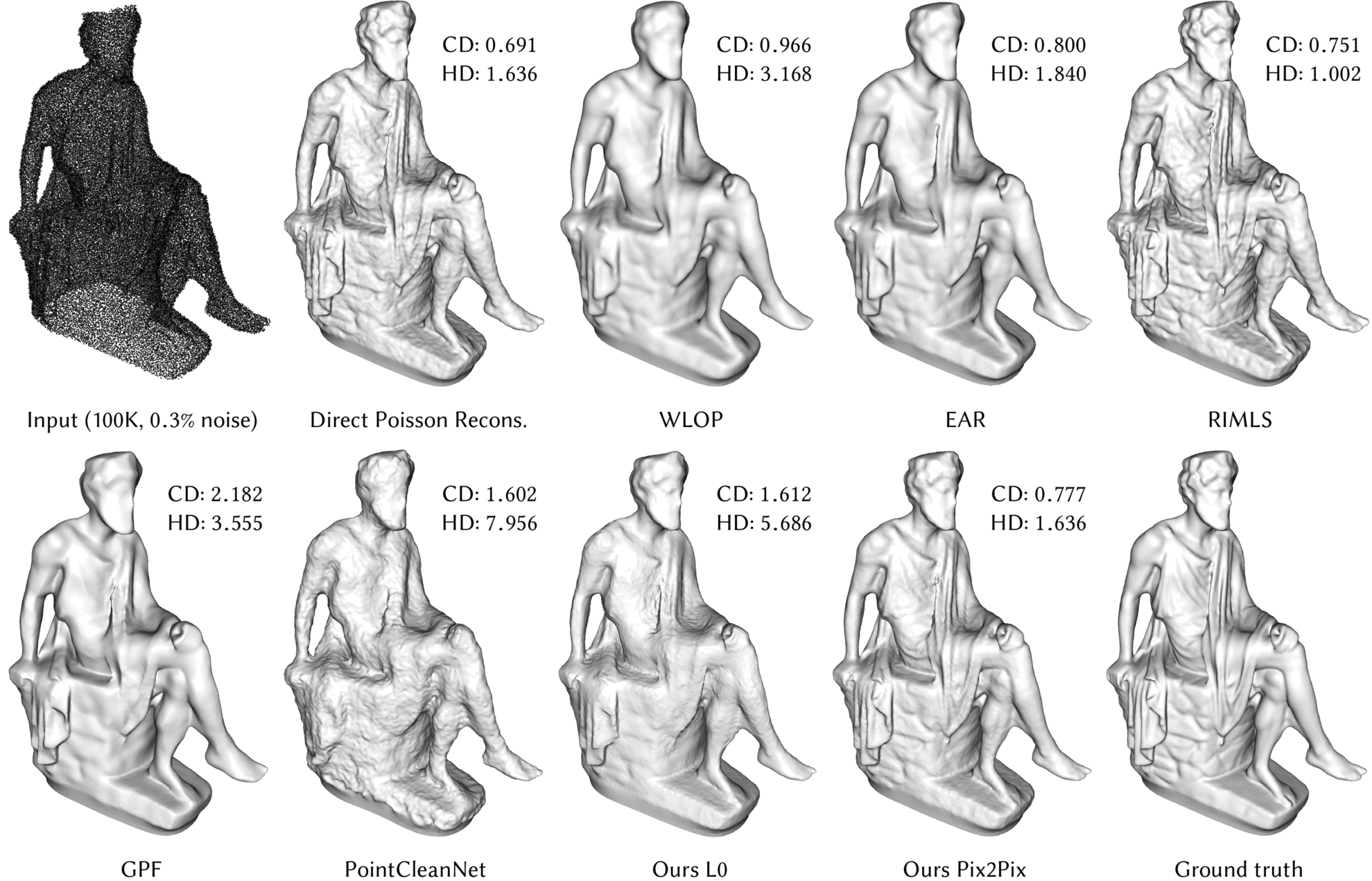}
	\caption{Quantitative and qualitative comparison of point cloud denoising with $ 0.3\% $ noise.  We report CD and HD scaled by $ 10^{-4} $ and $ 10^{-3} $. Despite some methods performing better with respect to quantitative evaluation, our result matches the ground truth closely in contrast to other methods.}\label{fig:seated}
\end{figure*}

To maximize the amount of hallucinated details, we train two models for 1.0\% and 0.3\% noise respectively.
\figref{fig:pix2pix_output} shows some examples of the input and output of the network.
Hallucinated delicate structures can be observed clearly in both noise levels.
Furthermore, even though our Pix2Pix model is not trained with view-consistency constraints, the hallucinated details remain mostly consistent across views.
In case small inconsistencies appear in regions where a large amount of high-frequency details are created, DSS is still able to transfer plausible details from the 2D to the 3D domain without visible artefacts, as shown in \figref{fig:seated}, thanks to simultaneous multi-view optimization.

\begin{figure}
\setlength{\tabcolsep}{1pt}
\begin{tabular}{ccccc}
\includegraphics[width=0.18\linewidth]{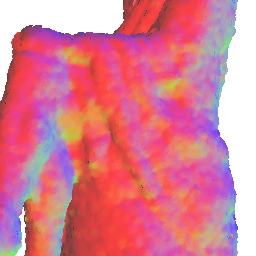}&
\includegraphics[width=0.18\linewidth]{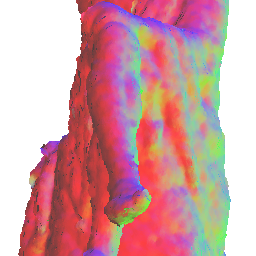}&
\includegraphics[width=0.18\linewidth]{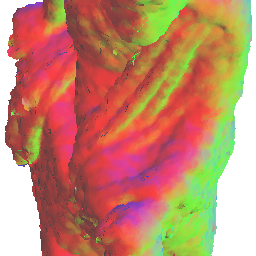}&
\includegraphics[width=0.18\linewidth]{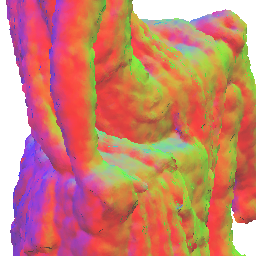}&
\includegraphics[width=0.18\linewidth]{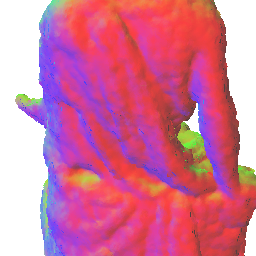}\\
\includegraphics[width=0.18\linewidth]{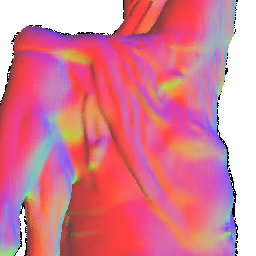}&
\includegraphics[width=0.18\linewidth]{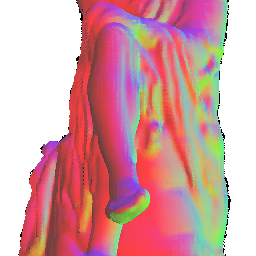}&
\includegraphics[width=0.18\linewidth]{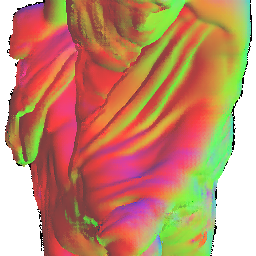}&
\includegraphics[width=0.18\linewidth]{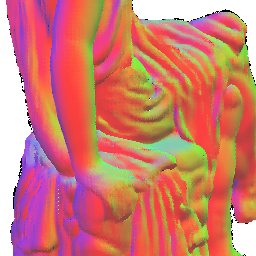}&
\includegraphics[width=0.18\linewidth]{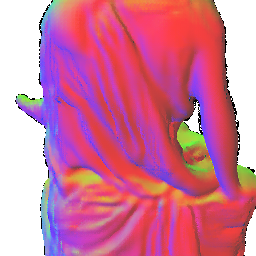}
\end{tabular}
\caption{Examples of multi-view Pix2Pix denoising on the same 3D model. As our Pix2Pix model processes each view independently, inconsistencies across different views might occur in the generated high-frequency details. In spite of that, DSS recovers plausible structures in the 3D shape (see \figref{fig:seated}) thanks to our simultaneous multi-view optimization.}\label{fig:multiview}
\end{figure}

\noindent \emph{Evaluation of DSS denoising.} 
We perform quantitative and qualitative comparison with state-of-the-art optimization-based methods WLOP~\cite{Huang2009wlop}, EAR~\cite{EAR2013}, RIMLS~\cite{oztireli2009feature} and GPF~\cite{lu2018gpf}, as well as a learning-based method, PointCleanNet~\cite{rakotosaona2019pointcleannet}, using the code provided by the authors. For quantitative comparison, we compute Chamfer distance (CD) and Hausdorff distance (HD) between the reconstructed and ground truth surface.

\begin{figure*}[h!]
 \centering
    \includegraphics[width=0.9\linewidth]{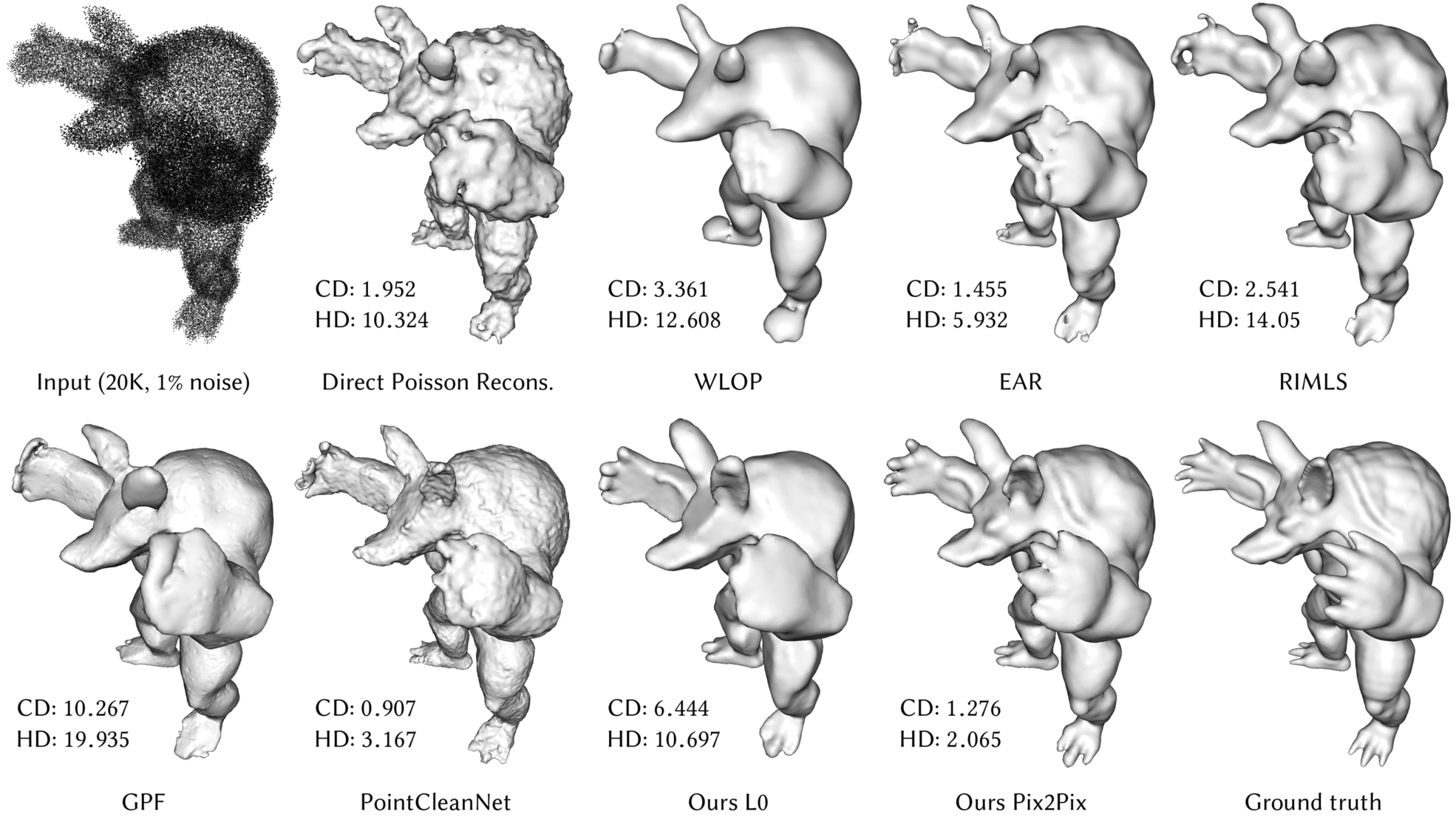}
	\caption{Quantitative and qualitative comparison of point cloud denoising. The Chamfer distance (CD) and Hausdorff distance (HD) scaled by $ 10^{-4} $ and $ 10^{-3} $. With respect to HD, our method outperforms its competitors, for CD only PointCleanNet can generate better, albeit noisy, results.}\label{fig:armadillo}
\end{figure*}

First, we compare the denoising performance on a relatively noisy (1\% noise) and sparse (20K points) input data, as shown in \figref{fig:armadillo}. Optimization-based methods can reconstruct a smooth surface but also smear the low-level details. The learning-based PointCleanNet can preserve some detailed structure, like the fingers of armadillo, but cannot remove all high-frequency noise. 
We test DSS with two image filters, i.e., the $L_0$ smoothing and the Pix2Pix model trained on data with 20K points and 1\% noise. 
$L_0$-DSS has a similar performance with the optimization-based method. Pix2Pix-DSS outperforms the other compared methods quantitatively and qualitatively.

Second, we evaluate on a relatively smooth (0.3\% noise) and dense (100K points) input data, as shown in \figref{fig:seated}. 
Optimization-based methods and  $L_0$-DSS produce high-accuracy reconstruction. 
PointCleanNet's result deteriorates significantly, due to generalizability issues which is common for direct learning-based methods. 
In contrast, the proposed image-to-geometry denoising method is inherently less sensitive to the characteristic of points sampling.
As a result, even though our Pix2Pix model is trained with 20K points, Pix2Pix-DSS reconstructs a clean surface, and at the same time shows abundant hallucinated details.

\begin{figure*}[h!]
	\includegraphics[width=0.9\linewidth]{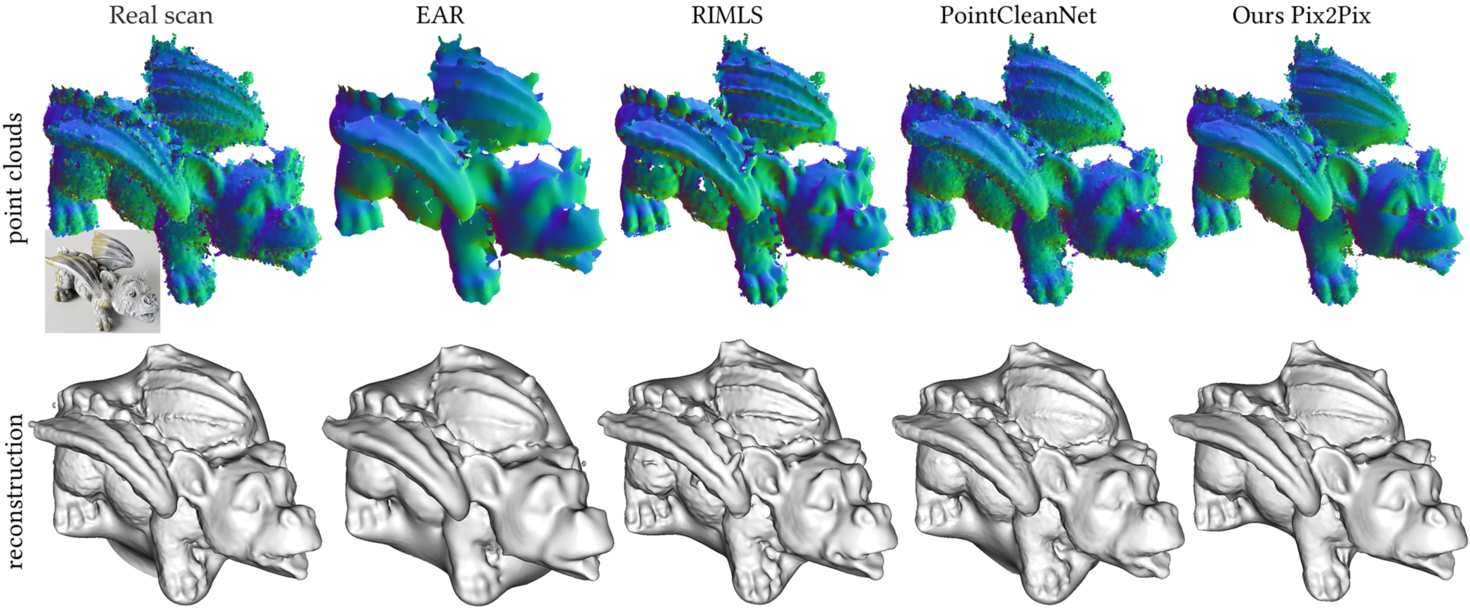}
	\caption{Qualitative comparison of point cloud denoising on a dragon model acquired using a hand-held scanner (without intermediate mesh representation). Our Pix2Pix-DSS outperforms the compared methods.}
	\label{fig:dragon}
\end{figure*}

Finally, we evaluate Pix2Pix-DSS using real scanned data.
We acquire a 3D scan of a dragon model by ourselves using a hand-held scanner and resample 50K points as input. 
We compare the point cloud cleaning performance of EAR, RIMLS, PointCleanNet and Ours as shown in \figref{fig:dragon}.
EAR outputs clean and smooth surfaces but tends to produce underwhelming geometry details.
RIMLS preserves sharp geometry features, but compared to our method, its result contains more low-frequency noise.
The output of PointCleanNet is notably noisier than other methods, while its reconstructed model falls between EAR and RIMLS in terms of detail preservation and surface smoothness.
In comparison, our method yields clean and smooth surfaces with rich geometry details.
\begin{table*}[h!]
	\small
	\begin{tabular}{ccccccccc}
		\toprule
		model & application & \makecell[c]{number of\\points} & \makecell[c]{total opt. steps\\ for position} & \makecell[c]{total opt. steps\\ for normal} & \makecell[c]{avg. forward\\time (ms)} & \makecell[c]{avg. backward\\time (ms)} & \makecell{total\\time (s)} & \makecell{GPU\\memory (MB)}\\
		\midrule
		\makecell{\figref{fig:compare_topology}} & shape deformation & 8003 &  200 & 120 & 19.3 & 79.9 & 336 & 1.7MB\\
		\makecell{\figref{fig:image_filtering}} & L0 surface filtering & 20000 &  8 & 152 & 42.8 & 164.6 & 665 & 1.8MB \\
		\makecell{\figref{fig:seated}} & denoising & 100000 & 8 & 152 & 258.1 & 680.2 & 1951 & 2.3MB \\
		\bottomrule
	\end{tabular}
	\caption{Runtime and GPU memory demand for exemplar models in different applications.
		The images are rendered with $ 256\times256 $ resolution and 12 views are used per optimization step.}\label{tab:speed}
\end{table*}

\subsection{Performance}
Our forward and backward rasterization passes are implemented in CUDA. 
We benchmark the runtime using an NVIDIA 1080Ti GPU with CUDA 10.0 and summarize the runtime as well as memory demand for all of the applications mentioned above on one exemplary model in Table~\ref{tab:speed}.
As before, models are rendered with $ 256\times256 $ resolution and 12 views are used per optimization step.

As a reference, for the teapot example, one optimization step in Paparazzi and Neural Mesh Renderer takes about 50ms and 160ms respectively, whereas it takes us 100ms (see the second row in Table~\ref{tab:speed}). 
However, since Paparazzi does not jointly optimize multiple-views, it requires more iterations for convergence.
In the L0-Smoothing example (see \figref{fig:papa_filter}), it takes 30 minutes and 30000 optimization steps to obtain the final result, whereas DSS needs 160 steps and 11 minutes for a similar result (see the third row in Table~\ref{tab:speed}).
\section{Conclusion and future works}

We showed how a high-quality splat based differentiable renderer could be developed in this paper. DSS inherits the flexibility of point-based representations, can propagate gradients to point positions and normals, and produces accurate geometries and topologies. These were possible due to the careful handling of gradients and regularization. We showcased a few applications of how such a renderer can be utilized for image-based geometry processing. In particular, combining DSS with contemporary deep neural network architectures yielded state-of-the-art results. 

There are a plethora of neural networks that provide excellent results on images for various applications such as stylization, segmentation, super-resolution, or finding correspondences, just to name a few. Developing DSS is the first step of transferring these techniques from image to geometry domain. Another fundamental application of DSS is in inverse rendering, where we try to infer scene-level information such as geometry, motion, materials, and lighting from images or video. We believe DSS will be instrumental in inferring dynamic scene geometries in multi-modal capture setups.

\begin{acks}
We would like to thank Federico Danieli for the insightful discussion, Phillipp Herholz for the timely feedack and Romann Weber for the video voice-over.
This work was supported in part by gifts from Adobe, Facebook and Snap, Inc.
\end{acks}

	\bibliographystyle{ACM-Reference-Format}
	\bibliography{bibliography}
	\appendix
	
\section{Parameter discussion}
\label{sec:parameters}
Here, we describe the effects of all hyper-parameters of our method.

\emph{Forward rendering.} The required hyper-parameters consist of $ \CC $ cutoff threshold, $ \TT $ merge threshold and $ \sigma_k $ (standard deviation). For all these parameters, we closely follow the default settings in the original EWA paper. \cite{zwicker2001surface}. For close camera views, the default $ \CC $ value is increased so that the splats are large enough to create hole-free renderings.

\emph{Backward rendering.} The cache size $ K $ used for logging points which are projected to each pixel is the only hyper-parameter. The larger $ K $ is, the more accurate $ \WW_{\xx, k} $ becomes, as more occluded points can be considered for the re-evaluation of \eqref{eq:sumI}. We find $ K=5 $ is sufficiently large for our experiments.

\emph{Regularization.} Bandwidth $ \DD $ and $ \Theta $ for computing weights in \eqref{eq:psi} and \eqref{eq:theta} are set as suggested in previous works \cite{Huang2009wlop,oztireli2009feature}. Specifically, $ \DD = 4\sqrt{D/N} $, where $ D $ is the diagonal length of the bounding box of the initial shape and $ N $ is the number of points; $ \Theta $ is set to $ \pi/3 $ to encourage a smooth surface under the presence of outliers.
For large-scale deformation, where the intermediate results can have more outliers, we set $ \DD $ of the projection term to a higher value, e.g. $ 0.1\sqrt{D} $, which helps to pull the outliers to the nearest surface.

\emph{Optimization.} The learning rate has a substantial impact on convergence. In our experiments, we set the learning rate for position and normal to 5 and 5000. These values generally work well for all applications. Higher learning rates cause the points to converge faster but increases the risk of causing the points to gather in clusters.
A more sophisticated optimization algorithm can be applied for a more robust optimization process, but it is out of the scope of this paper.
A sufficient number of views per optimization step is key to a good result in the ill-posed 2D-to-3D formulation.
Twelve camera views are used in all our experiments, while with 8 or fewer views results start to degenerate.
The number of steps for points and normals update, $ T_\pp $ and $ T_\nn $, differ for each application. 
In general, for large topology changes, we set $ T_\pp > T_\nn $, where typically $ T_\pp = 25 $ and $ T_\nn = 15 $, while for local geometry processing $ T_\nn > T_\pp $ with $ T_\nn = 19 $ and $ T_\nn = 1 $.
Finally, we find the loss weights for image loss $ \LL_\Image $, projection regularization $ \LL_{p} $ and repulsion regularization $ \LL_{r} $, by ensuring the magnitude of per point gradient from $ \LL_{p} $ and $ \LL_{r} $ is around $ 1\% $ of that from $ \LL_\Image $.
If the repulsion weight is too large, e.g. $ \gamma_r > 0.1 $, points can be repelled far off the surface, while if the projection weight is too large, e.g. $ \gamma_p > 0.1 $, points will be forced to stay on a local surface, making it difficult for topology changes.

\section{Denoising Pix2Pix}\label{sec:pix2pix}
Our model is based on Pix2Pix~\cite{isola2017image} that consist of a generator and a discriminator. For the generator, we experimented  with U-Net \cite{ronneberger2015uet} and ResNet \cite{He_2016_CVPR}, and find ResNet performs slightly better in our task, which we use for all experiments.  
That is, the generator has a 2-stride convolution and a 2-stride up-convolution for both the encoder and decoder networks and 9 residual blocks in-between. 
The discriminator follows the architecture as: C64-C128-C256-C512-C1, where LSGAN~\cite{mao2017least} is used. 
To deal with checkerboard artifacts, we use pixel-wise normalization in the generator and add a convolutional layer after each deconvolutional layer in the discriminator~\cite{karras2017progressive}. 
Furthermore, we remove the \texttt{tanh} activation in the final layer in order to obtain unbounded pixel values
We use the default parameters of the Pix2Pix pytorch implementation provided by the authors, and ADAM optimizer  ($lr = 0.0002, \beta_1 = 0.5, \beta_2 = 0.999$) .
Xavier~\cite{glorot2010understanding} is used for weights initialization. We train our models for about two days on an NVIDIA 1080Ti GPU.

To synthesize training data for the Pix2Pix denoising network, we use the training set of the Sketchfab dataset~\cite{yifan2018patch}, which consist of 91 high-resolution 3D models.
We use Poisson-disk sampling~\cite{corsini2012efficient} implemented in Meshlab~\cite{Cignoni2008MeshLabAO} to sample 20K points per model as reference points, and create noisy input points by adding white Gaussian noise, then we compute the PCA normal~\cite{Hoppe1992} for both the reference and input points.
We generate training data by rendering a total of 149240 pairs of images from the noisy and clean models using DSS, from a variety of viewpoints and distances. 
We use point light and diffuse shading. 
While using sophisticated lighting, non-uniform albedo and specular shading can provide useful cues for estimating global information such as lighting and camera positions, we find the glossy effects pose unnecessary difficulties for the network to infer local geometric structure.

%

\end{document}